\documentclass[a4paper,11pt]{article}
\pdfoutput=1

\usepackage{jheppub}
\usepackage{graphicx}
\usepackage{amsmath}
\usepackage{mathrsfs}
\usepackage{amssymb}
\usepackage{xspace}
\usepackage{xcolor}
\usepackage{snapshot}

\newcommand{\refcite}[1]{ref.~\cite{#1}}
\newcommand{\refscite}[1]{refs.~\cite{#1}}
\newcommand{\eq}[1]{eq.~\eqref{eq:#1}}

\renewcommand{\sec}[1]{sec.~\ref{sec:#1}}

\newcommand{\subsec}[1]{sec.~\ref{subsec:#1}}

\newcommand{\app}[1]{app.~\ref{app:#1}}
\newcommand{\fig}[1]{fig.~\ref{fig:#1}}
\newcommand{\Fig}[1]{Fig.~\ref{fig:#1}}
\newcommand{\figs}[2]{figs.~\ref{fig:#1} and \ref{fig:#2}}
\newcommand{\tab}[1]{table~\ref{tab:#1}}

\newcommand{\nn}{\nonumber}

\newcommand{\ord}[1]{\mathcal{O}(#1)}

\newcommand{\df}{\mathrm{d}}

\newcommand{\al}{\alpha}
\newcommand{\bt}{\beta}
\newcommand{\ga}{\gamma}
\newcommand{\Ga}{\Gamma}
\newcommand{\de}{\delta}

\newcommand{\si}{\sigma}

\let\oldvec\vec
\renewcommand*\vec[1]{\oldvec{\kern0pt #1}}

\newcommand{\cL}{\mathcal{L}}
\newcommand{\cS}{\mathscr{S}}

\newcommand{\cusp}{\mathrm{cusp}}

\newcommand{\cut}{\mathrm{cut}}

\newcommand{\MSbar}{$\overline{\text{MS}}$\xspace}

\newcommand{\Pythia}{\textsc{Pythia}\xspace}
\newcommand{\FastJet}{\textsc{FastJet}\xspace}
\newcommand{\Herwig}{\textsc{Herwig}\xspace}
\newcommand{\Rambo}{\textsc{Rambo}\xspace}

\newcommand{\GeV}{\,\mathrm{GeV}}

\allowdisplaybreaks[2]

\title{\boldmath How much joint resummation do we need?}

\author[a,b]{Gillian Lustermans,}
\author[a,b]{Andreas Papaefstathiou,}
\author[a,b]{and Wouter J.~Waalewijn}

\affiliation[a]{Institute for Theoretical Physics Amsterdam and Delta Institute for Theoretical Physics, University of Amsterdam, Science Park 904, 1098 XH Amsterdam, The Netherlands}
\affiliation[b]{Nikhef, Theory Group, Science Park 105, 1098 XG, Amsterdam, The Netherlands}

\emailAdd{g.h.h.lustermans@uva.nl}
\emailAdd{a.papaefstathiou@uva.nl}
\emailAdd{w.j.waalewijn@uva.nl}

\abstract{
Large logarithms that arise in cross sections due to the collinear and soft singularities of QCD are traditionally treated using parton showers or analytic resummation. Parton showers provide a fully-differential description of an event but are challenging to extend beyond leading logarithmic accuracy. On the other hand, resummation calculations can achieve higher logarithmic accuracy but often for only a single observable. Recently, there have been many resummation calculations that jointly resum multiple logarithms. Here we investigate the benefits and limitations of joint resummation in a case study, focussing on the family of $e^+e^-$ event shapes called angularities. We calculate the cross section differential in $n$ angularities at next-to-leading logarithmic accuracy. We investigate whether reweighing a flat phase-space generator to this resummed prediction, or the corresponding distributions from \Herwig and \Pythia, leads to improved predictions for other angularities. We find an order of magnitude improvement for $n = 2$ over $n=1$, highlighting the benefit of joint resummation, but diminishing returns for larger values of $n$.}
\preprint{\vbox{
\hbox{Nikhef 19-032}}}

\begin{document}

\maketitle

%%%%%%%%%%%%%%%%%%%%%%%%%%%%%%%%%%%%%%%%%%%%%%%%%%%%%%%%%%%%%%%%%%%%%%%%%%%%%%%%
\section{Introduction}
\label{sec:intro}
%%%%%%%%%%%%%%%%%%%%%%%%%%%%%%%%%%%%%%%%%%%%%%%%%%%%%%%%%%%%%%%%%%%%%%%%%%%%%%%%

Measurements at colliders often impose restrictions on QCD radiation through e.g.~jet vetoes, transverse momentum measurements, or production at threshold. One particular example, that will play a prominent role in this paper, is the family of event shapes for $e^+e^-$ collisions called angularities~\cite{Berger:2003iw}. These are defined as
%%%
\begin{align} \label{eq:ang_def}
  e_\al = \frac{2}{Q} \sum_i E_i \Bigl[\sin \Bigl(\frac{\theta_i}{2}\Bigr)\Bigr]^\al
\,,\end{align}
%%%
where $Q$ is the center-of-mass energy, and the sum runs over all final-state particles $i$ with energy $E_i$ and angle $\theta_i$ with respect to a chosen axis.
The cross section integrated over $e_\al$ up to some cut-off $e_\al^{\rm cut}$, known as the cumulative cross section, contains a series of logarithms $L = \log_{10} e_\al^{\rm cut}$ at each order in perturbation theory, schematically
%%%
\begin{align}
  \int_0^{e_\al^{\rm cut}}\,\df e_\al\, \frac{\df \si}{\df e_\al} &= \si_0 \bigl[1 & \text{LO}
  \nn \\[-1.5ex] & \quad
  +  \al_s (c_{12} L^2 + c_{11} L + c_{10}) & \text{NLO}
  \nn \\ & \quad
  +  \al_s^2 (c_{24} L^4 + c_{23} L^3 + c_{22} L^2 + c_{21} L + C_{20})  +  \ord{e_\al^{\rm cut}, \al_s^3}\bigr]\,. & \text{NNLO}
  \nn \\ & \qquad\qquad \text{LL}  \qquad \text{NLL} \qquad \text{NNLL}
\end{align}
%%%
Rows correspond to different orders of fixed-order perturbation theory, denoted by leading order (LO), next-to-leading order (NLO), etc. For $e_\al^{\rm cut} \ll 1$, the $\al_s$ expansion breaks down, because we cannot treat $L \sim 1$. In this case we want to sum columns, which correspond to different orders of resummed perturbation theory, called leading logarithmic (LL) order, next-to-leading logarithmic (NLL) order, etc. To be precise, we will treat $\al_s L \sim 1$, which corresponds to resumming logarithms ``in the exponent".

This resummation can be carried out either by using a Monte Carlo parton shower, such as \Herwig~\cite{Bahr:2008pv, Bellm:2015jjp, Bellm:2017bvx} or \Pythia~\cite{Sjostrand:2014zea}, or through analytical methods. The advantage of parton showers is that they allow for any measurement on the fully exclusive final state. However, their formal accuracy is limited to LL order in the large $N_c$ limit, and it is challenging to systematically go beyond this order. For a recent discussion of the logarithmic accuracy of the parton shower, see refs.~\cite{Hoeche:2017jsi,Dasgupta:2018nvj,Bewick:2019rbu}. Some recent improvements in parton showers are the inclusion of higher-order splitting functions~\cite{Jadach:2016zgk,Li:2016yez,Hoche:2017iem,Hoche:2017hno}, corrections to the large $N_c$ limit~\cite{Nagy:2015hwa,Isaacson:2018zdi,Platzer:2018pmd}, spin correlations~\cite{Richardson:2018pvo}, and the simultaneous treatment of small $x$ and collinear and soft logarithms~\cite{Andersen:2017sht}.

On the other hand, analytic resummation calculations are able to achieve a much higher precision. As an example, in the case of angularities, predictions at NNLL+NNLO accuracy are available~\cite{Banfi:2018mcq,Bell:2018gce}. Methods for analytic resummation include the CSS formalism~\cite{Collins:1985ue,Collins:1988ig,Collins:1989gx}, those based on the coherent-branching formalism~\cite{Banfi:2001bz, Banfi:2004yd, Banfi:2014sua}, and those using renormalization group evolution in effective field theories of QCD, such as Soft-Collinear Effective Theory (SCET)~\cite{Bauer:2000ew, Bauer:2000yr, Bauer:2001ct, Bauer:2001yt}. While many of these calculations have focussed on the resummation of a single logarithmic series, there has recently been a significant effort to jointly resum multiple logarithms. This includes the joint resummation of logarithms due to threshold production and transverse momentum~\cite{Li:1998is, Laenen:2000ij, Kulesza:2002rh, Kulesza:2003wn, Lustermans:2016nvk, Marzani:2016smx, Muselli:2017bad}, threshold and small $x$~\cite{Ball:2013bra,Bonvini:2018ixe}, transverse momentum and small $x$~\cite{Marzani:2015oyb}, transverse momentum and beam thrust~\cite{Procura:2014cba,Lustermans:2019plv}, jet mass and dijet invariant mass~\cite{Bauer:2011uc, Pietrulewicz:2016nwo}, two angularities~\cite{Larkoski:2014tva, Procura:2018zpn}, jet veto and jet radius~\cite{Banfi:2015pju}, jet mass and jet radius~\cite{Kolodrubetz:2016dzb}, jet vetoes and jet rapidity~\cite{Hornig:2017pud, Michel:2018hui}, and threshold and jet radius~\cite{Liu:2017pbb, Liu:2018ktv}. 

In principle, one could imagine the simultaneous resummation of ever more observables, which would lead to an increasingly precise parton shower. This is important for the burgeoning field of Machine Learning in jet substructure, see ref.~\cite{Larkoski:2017jix} for a review. Here, samples from Monte Carlo parton showers are often employed, thus raising the question to what extent discrimination is based on features of physics or of the Monte Carlo. There has also been some work on approaches that do not require labeled samples though, see e.g.~refs.~\cite{Metodiev:2017vrx,Cohen:2017exh,Metodiev:2018ftz,Andreassen:2018apy}. We were inspired by ref.~\cite{Datta:2017rhs}, which investigates how sensitive Machine Learning is to details of the final state, studying the discrimination of jets from boosted, hadronic decays of $Z$ bosons from jets initiated by QCD processes. By using a complete basis of observables that probe the $N$-body final state and increasing $N$, they find that the discrimination saturates at $N=4$. 

We are interested in asking how much resummation is needed to reliably describe the jet, focussing on QCD-initiated jets. For simplicity, we restrict ourselves to observables that are azimuthally symmetric, for which the angularities form a basis. Specifically, we consider a set of $N$ angularities $e_{\alpha_i}$, with $i=1, \dots, N$, derive the resummed prediction for the cross section differential in a subset of angularities denoted by $I =\{i_1, \dots, i_n\}$ at NLL accuracy, and investigate the degree to which any of the other angularities $e_{\al_j}$, with $j \notin I$, can be predicted. To this end, we generate events from flat\footnote{``Flat'' here implies that there are no preferred points in phase space, so that, up to four-momentum conservation, each point is assigned the same probability.} $k$-body phase space, using \Rambo~\cite{Kleiss:1985gy} ``on diet''~\cite{Platzer:2013esa}. We reweigh these events to match the multi-differential cross sections from the input, and then calculate the distributions for other angularities from these reweighed events. As an alternative to analytic resummation, we also use the cross section differential in $n$ angularities from  \Herwig or \Pythia as basis for reweighing. The dependence of this procedure on the number $n$ of angularities that have been jointly resummed is investigated and the optimal subset $I$ of angularities to be used as input for any given $n$ is determined.

Our main conclusion is that there is an order of magnitude improvement for reweighing with $n=2$ angularities over $n=1$, but that the advantage quickly diminishes for larger values of $n$.
We have investigated the dependence on the input (\Herwig, \Pythia, analytic), number of particles $k$ of the flat phase space, center-of-mass energy $Q$, and set of angularities under consideration. None of these lead to a qualitatively different behavior.

The outline of this paper is as follows: We start in \sec{reweigh} by describing our setup and discussing in detail how we perform the reweighing, determine the optimal input set of angularities $I$, and estimate the statistical uncertainty. In \sec{jointresum}, we present our analytic calculation of the cross section differential in $n$ angularities, limiting most of the discussion to NLL accuracy. Our main results are presented in \sec{results}, with additional plots relegated to \app{extra}. We conclude in \sec{conc}.

%%%%%%%%%%%%%%%%%%%%%%%%%%%%%%%%%%%%%%%%%%%%%%%%%%%%%%%%%%%%%%%%%%%%%%%%%%%%%%%%
\section{Optimal reweighing procedure}
\label{sec:reweigh}
%%%%%%%%%%%%%%%%%%%%%%%%%%%%%%%%%%%%%%%%%%%%%%%%%%%%%%%%%%%%%%%%%%%%%%%%%%%%%%%%

We start this section by presenting our setup, discussing the observables in more detail. We then describe the reweighing procedure of flat $k$-body phase space with resummed predictions for $n$ angularities, as well as the determination of the optimal set of angularities to be used as input, including the treatment of statistical uncertainties.

%===============================================================================
\subsection{Setup}
\label{subsec:setup}
%===============================================================================

We use the process $e^+e^- \to \text{dijets}$ as a case-study for our procedure. The final state of the collision is clustered into two jets using the exclusive $k_T$~\cite{Catani:1991hj} jet algorithm with the Winner-Take-All (WTA) recombination scheme~\cite{Salam:WTAUnpublished,Bertolini:2013iqa}. Following ref.~\cite{Larkoski:2014uqa}, we modify the original definition of the angularities in ref.~\cite{Berger:2003iw} for large angles $\theta_i \sim 1$, as described by \eq{ang_def}. The angle $\theta_i$ of each particle contributing to an angularity is measured with respect to the WTA axis of the corresponding jet, and the angularities $e_\al$ that we consider are the sum of the angularities $e_\al^{\text{Jet }J}$ of the individual jets, i.e. $e_\al = e_\al^{\text{Jet 1}} + e_\al^{\text{Jet 2}}$. The variable $\ell_{\alpha_i} \equiv\log_{10} e_{\alpha_i}$ is used instead of the angularities themselves, which is more natural since the angularity distributions are peaked at small values of $e_{\alpha_i}$.\footnote{For brevity, we will also refer to the observables $\ell_{\alpha_i}$ as ``angularities".} 

The various distributions of $\ell_{\alpha_i}$ are constructed in 16 bins on the interval $\ell_{\alpha_i} \in [-4, 0]$. We have verified that the chosen binning does not alter the conclusions of our study. The binned distributions are constructed either by using analytic predictions (described in \sec{jointresum}) or through $10^6$ parton-level events generated by the \Herwig 7 or \Pythia 8 general-purpose Monte Carlos (described in \subsec{optimal_reweighing}).

%===============================================================================
\subsection{Optimal reweighing}
\label{subsec:optimal_reweighing}
%===============================================================================

To obtain the binned distributions for flat massless $k$-body phase space,
%%%
\begin{align}
       \int \prod_{i=1}^k \frac{\df^3 \vec p_i}{2|\vec p_i|}\, \delta^4\biggl(\sum_{j=1}^k p_j^\mu - Q^\mu\biggr)\,,
\end{align}
%%%
with $Q^\mu = (Q,0,0,0)$ in the center-of-mass frame, we employ the \Rambo technique of refs.~\cite{Kleiss:1985gy, Platzer:2013esa}, with a slight modification that improves sampling in the collinear and soft regions through weighted events. 
In particular, we perform a transformation that distributes the first random number of ``Algorithm 1'' in ref.~\cite{Platzer:2013esa} logarithmically, with the weight of this event given by the Jacobian.
This ensures that the phase space is sampled sufficiently to obtain statistically reliable predictions for small values of the angularities that we consider.

We start by reweighing the flat phase space result for $n+1$ angularities by the cross section differential in $n$ angularities $\ell_{\al_i}$ with $i \in I = \{i_1,\dots, i_n\}$. By integration\footnote{In our case, the integration is approximated by a sum over bins.} over the $n$ angularities $\ell_{\al_i}$, a reweighed cross section $\df \sigma_{\rm reweigh}/\df \ell_{\al_j}$ differential in the remaining angularity $\ell_{\al_j}$ with $j \notin I$ is determined, i.e.

%%%
 \begin{align}\label{eq:rw}
       \frac{\df \sigma_{\rm reweigh}}{\df \ell_{\al_j}} = 
       \int \prod_{i \in I} \df \ell_{\alpha_i}\,
       \frac{\df \sigma_{\rm flat}}{\df \ell_{\al_j} \prod_{i \in I} \df \ell_{\alpha_i}}  \times \frac{\df \sigma_{\rm resum}}{\prod_{i \in I} \df \ell_{\alpha_i}} \Big /  \frac{\df \sigma_{\rm flat}}{\prod_{i \in I} \df \ell_{\alpha_i}}\,,
\end{align}
%%%    
appropriately applied to the binned distributions.
Comparisons between the resulting reweighed cross sections and direct determinations of equivalent cross sections can be found in \figs{single}{singleresum}.

We define a goodness-of-fit measure for the reweighed angularity distribution by comparing to the resummed distribution for $\ell_{\al_j}$ 
%%%
\begin{align} \label{eq:chi2_beta}
    \chi_{\alpha_j}^2 = \int\! \df \ell_{\al_j}\,  \Big|\frac{\df \sigma_{\rm reweigh}}{\df \ell_{\alpha_j}} - \frac{\df \sigma_{\rm resum}}{\df \ell_{\alpha_j}}\Big|^2
\,.\end{align}
%%%
To find the optimal set $I$ when reweighing with $n$ angularities, we introduce a global goodness-of-fit variable, defined as the average $\chi_{\alpha_j}^2$ of the angularities $j \notin I$,
%%%
\begin{align} \label{eq:chi2_total}
  \chi^2 = \frac{1}{N-n} \sum_{j \notin I} \chi_{\alpha_j}^2\,.
\end{align}
%%%
Here $N-n$ is the number of angularities not used as input. 

Up to $n=3$, we search for the global minimum of $\chi^2$ as a function of the set of reweighed angularities $I$, denoted by $\chi^2_{\rm min}$. We will refer to this as the optimal set of angularities in each case, and the corresponding value for $\chi_{\alpha_i}^2$ will be denoted by $\chi_{\alpha_i, {\rm min}}^2$. For $n = 4$ and $n = 5$, we start from the optimal $I$ for $n-1$ angularities and iteratively determine the optimal additional angularity to add to $I$. We have verified that the result of this iterative method is already very close to the global minimum of $\chi^2$ for $n=3$. 

The value of the global minimum $\chi^2_{\rm min}$ is rather sensitive to statistical fluctuations, owing to the finite size of the Monte Carlo samples distributed over $16^{n+1}$ bins.\footnote{The $n+1$-dimensional distributions are required to perform the projection in \eq{rw}.} Furthermore, it is not expected to follow a Gaussian distribution. To obtain an estimate of the statistical uncertainty, we perform the reweighing procedure over 11 replicas of the event samples. The median of $\chi^2_{\rm min}$ is taken as our central prediction, and the spread of the 7 most central replicas as a reasonable approximation for the spread corresponding to roughly one standard deviation. Results for $\chi^2_{\rm min}$ as a function of $n$ are depicted in \fig{global} and in \app{robustness}.

%%%%%%%%%%%%%%%%%%%%%%%%%%%%%%%%%%%%%%%%%%%%%%%%%%%%%%%%%%%%%%%%%%%%%%%%%%%%%%%%
\section{Joint resummation of $n$ angularities at NLL}
\label{sec:jointresum}
%%%%%%%%%%%%%%%%%%%%%%%%%%%%%%%%%%%%%%%%%%%%%%%%%%%%%%%%%%%%%%%%%%%%%%%%%%%%%%%%

In this section we present our framework for performing the joint resummation of $n$ angularities. We start by drawing the Lund planes in \subsec{Lund}, from which the leading-logarithmic resummation immediately follows. These diagrams also allow us to identify the  modes in SCET, for which the factorization formulae are presented in \subsec{fact}. From the renormalization group equations for these factorization formulae we obtain the resummed cross section at NLL accuracy in \subsec{resum}. We describe the matching of the different factorization formulae for different regions of phase space in \subsec{match}, for which we are guided by the size of the power corrections, estimated in \subsec{pc}.

%===============================================================================
\subsection{Lund diagrams and phase-space boundaries}
\label{subsec:Lund}
%===============================================================================

\begin{figure}
\centering
\includegraphics[width=.4\textwidth]{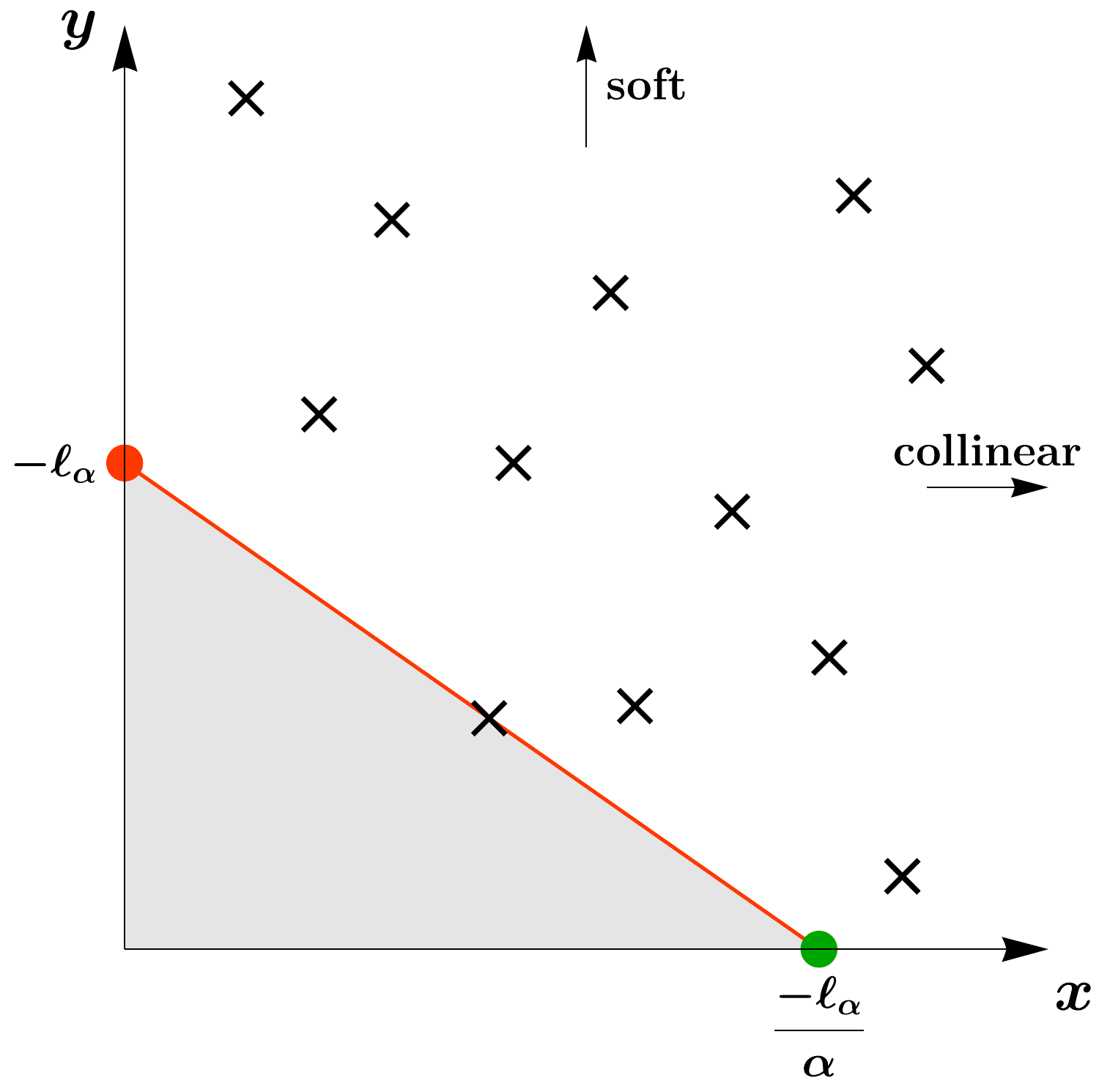}%
\caption{Illustration of the Lund plane with $x = \log_{10}(1/\theta)$ and $y = \log_{10}(1/z)$. The crosses represent emissions and the red line the measurement of an angularity $\ell_\al$, set by the dominant emission. There can be no emissions in the shaded area, while emissions above this only contribute at higher order. The green and orange dot denote the collinear and soft mode respectively.}
\label{fig:LundPlane1}
\end{figure}

In the collinear limit, the probability $P$ of a particle emitting real radiation can be characterized by the momentum fraction $z$ and angle $\theta$ of the radiated particle $j$ with respect to its emitter $i$, 
%%%
\begin{align}
  \frac{\df P_{i \to j}}{\df \theta\, \df z} = \frac{\al_s}{\pi}\, \frac{P_{i \to j}(z)}{\theta} 
\,,\end{align}
%%%
At LL accuracy, only the $\sim 1/z$ term of the splitting function $P_{i \to j}(z)$ is kept, so these emissions are uniformly distributed in $x \equiv \log_{10}(1/\theta)$ and $y \equiv \log_{10}(1/z)$. The Lund plane~\cite{Andersson:1988gp} spanned by these variables is shown in \fig{LundPlane1}, with some emissions indicated by crosses. 

\begin{figure}
\centering
\hspace*{-.9cm}%
\includegraphics[height=.33\textwidth]{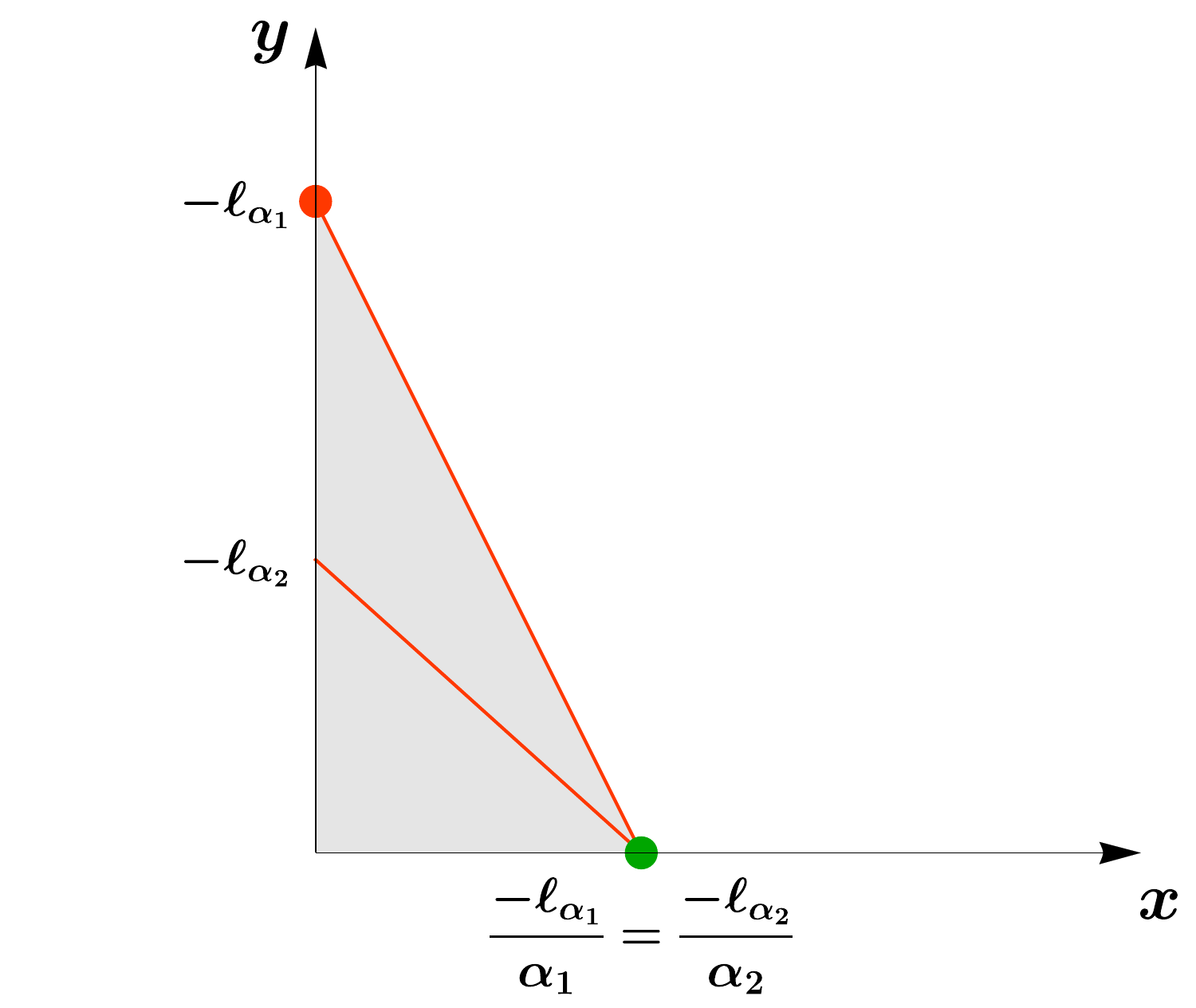}%
\hspace*{-.9cm}%
\includegraphics[height=.33\textwidth]{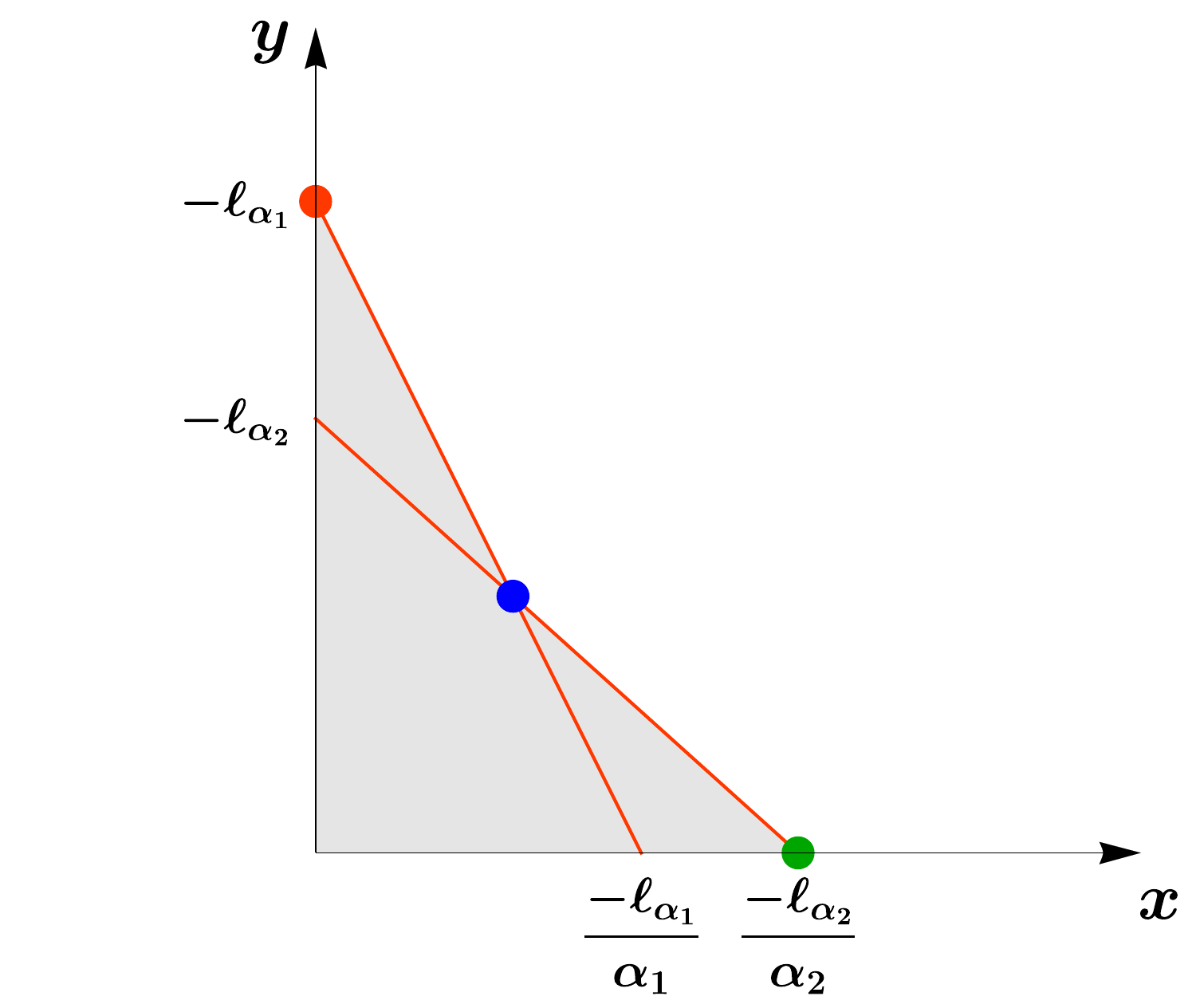}%
\hspace*{-.9cm}%
\includegraphics[height=.33\textwidth]{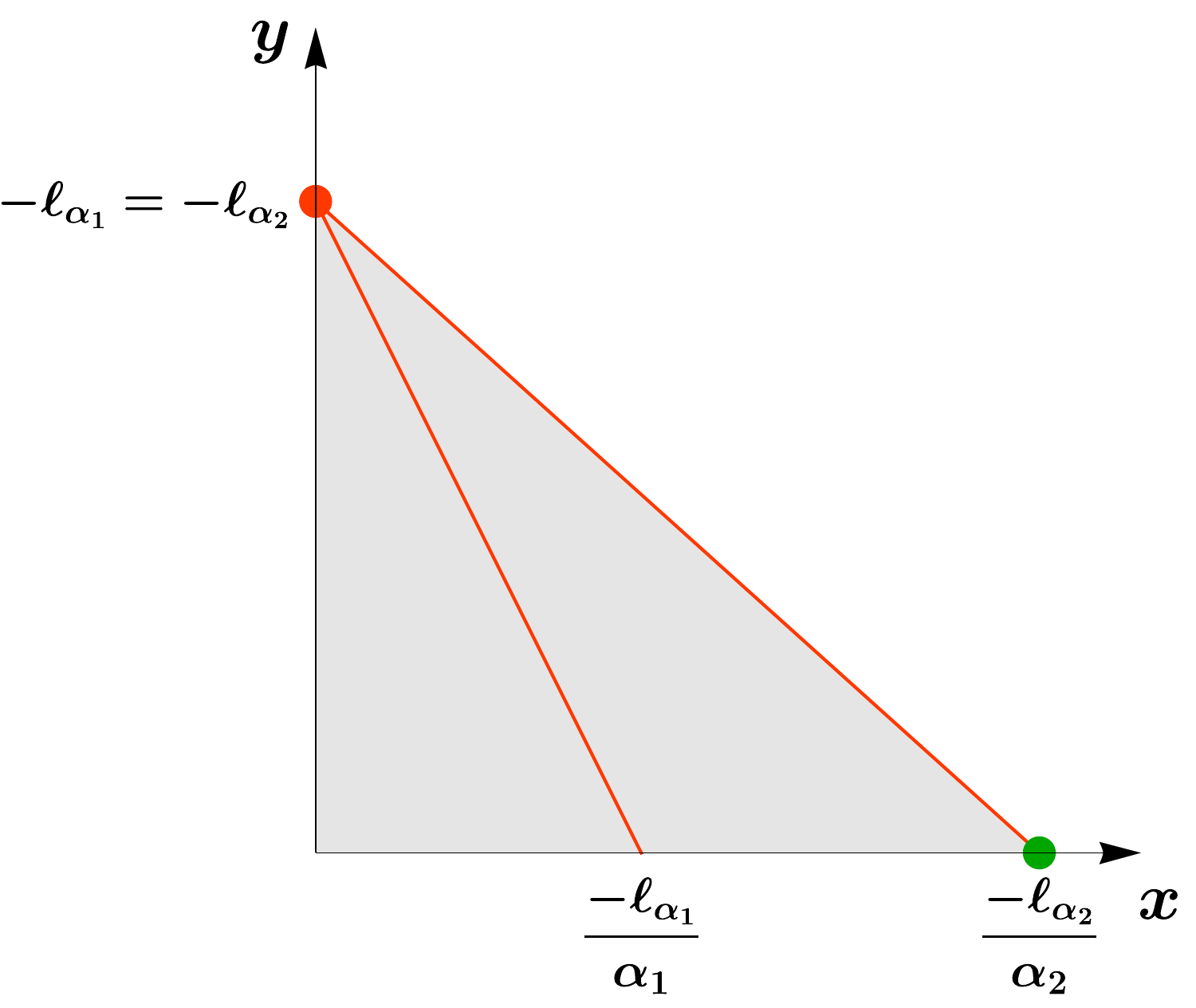}%
\caption{The measurement of two angularities $\ell_{\al_1}$ and $\ell_{\al_2}$ represented in the Lund plane. Each panel describes a distinct region of phase space. The left and right panels involve only collinear (green) and soft (orange) modes, while the center panel contains an additional collinear-soft (blue) mode. Emissions in the shaded region are vetoed.}
\label{fig:LundPlane2}
\end{figure}

By identifying $2E_i/Q \to z_i$ and $\theta_i/2 \to \theta$ (the factor of $1/2$ is purely for convenience and doesn't affect the leading logarithms), the angularity in \eq{ang_def}  for a single emission with small $z$ and $\theta$ can be written as
%%%
\begin{align}
\ell_\al = - y - \al x
\,,\end{align}
%%%
corresponding to a straight line in the Lund plane with slope $-\al$. Due to their uniform distribution in the (logarithmically-spaced) Lund plane, a single emission will dominate the measurement at this accuracy, as indicated by the red line in \fig{LundPlane1}. All emissions above and to the right of this line are more soft or collinear and only enter beyond LL accuracy. There are no emissions below the line, otherwise these would be dominant. The shaded area under the line corresponds to the Sudakov factor describing the no-emission probability, and can be used to calculate the cumulative cross section,
%%%
\begin{align}
  \si (e_\al < e_\al^{\rm cut}) = \hat{\si}_0\,\exp\Bigl(- \frac{4\al_s C_i}{\pi} \times [\text{gray area}] \times \ln^2 10\Bigr)
\,,\end{align}
%%%
where $\hat{\si}_0$ is the Born cross section and $C_i = C_F$ ($C_A$) for quark (gluon) jets.
Interestingly, the relevant degrees of freedom in SCET correspond to the points that describe the edges of the shaded region. For the measurement of a single angularity these are indicated by the orange and green dot in \fig{LundPlane1}, and correspond to soft and collinear modes respectively. The parametric scaling of the momenta of these modes is most conveniently expressed in terms of lightcone coordinates defined through
%%%
\begin{align}
  p^\mu = \frac{\bar{n}^\mu}{2} p^+ + \frac{n^\mu}{2} p^- +  p_\perp^\mu \equiv (p^+,p^-,\vec{p}_\perp)
\,.\end{align}
%%%
Indicating the momenta of the soft, $n$-collinear and $\bar{n}$-collinear modes by $p_s^\mu$, $p_n^\mu$ and $p_{\bar{n}}^\mu$ respectively, their scaling is found to be~\cite{Hornig:2009vb}
%%%
\begin{alignat}{2} \label{eq:modes_scaling_angularities}
  p_n^\mu &\sim Q\bigl(e_\al^{2/\al},1,e_\al^{1/\al}\bigr)\,, \nn\\
  p_{\bar{n}}^\mu &\sim Q\bigl(1, e_\al^{2/\al},e_\al^{1/\al}\bigr)\,, \nn\\
  p_s^\mu &\sim Q\bigl(e_\al,e_\al,e_\al\bigr)
\,.\end{alignat}
%%%
When the simultaneous measurement of two angularities $\ell_{\al_1}$ and $\ell_{\al_2}$ is considered, the straight lines describing the variables in the Lund plane have to cross one another at some point to ensure that the cross section depends on both measurements. Assuming the hierarchy $\al_1 > \al_2$ for definiteness, three distinct cases can be distinguished, as shown in \fig{LundPlane2}.
The boundaries of these three regions of phase space for two angularities are 
%%%
\begin{align} \label{eq:regimes_2_angularities}
\text{Regime 1:} \qquad -\ell_{\al_2} < -\ell_{\al_1} \quad \text{and} \quad \frac{-\ell_{\al_1}}{\al_1} = \frac{-\ell_{\al_2}}{\al_2}\,, \nn\\
\text{Regime 2:}  \qquad -\ell_{\al_2} < -\ell_{\al_1} \quad \text{and} \quad \frac{-\ell_{\al_1}}{\al_1} < \frac{-\ell_{\al_2}}{\al_2}\,, \nn\\
\text{Regime 3:} \qquad -\ell_{\al_2} = -\ell_{\al_1} \quad \text{and} \quad \frac{-\ell_{\al_1}}{\al_1} < \frac{-\ell_{\al_2}}{\al_2}\,,
\end{align}
%%%
which agree with the regions of phase space identified in \refscite{Larkoski:2014tva,Procura:2014cba,Procura:2018zpn}. In all three cases there are soft (orange) and collinear (green) degrees of freedom. The intermediate regime 2 has an additional collinear-soft mode (blue),  which contributes to both measurements since it lies on the intersection of both lines.

\begin{figure}
\centering
\includegraphics[width=.4\textwidth]{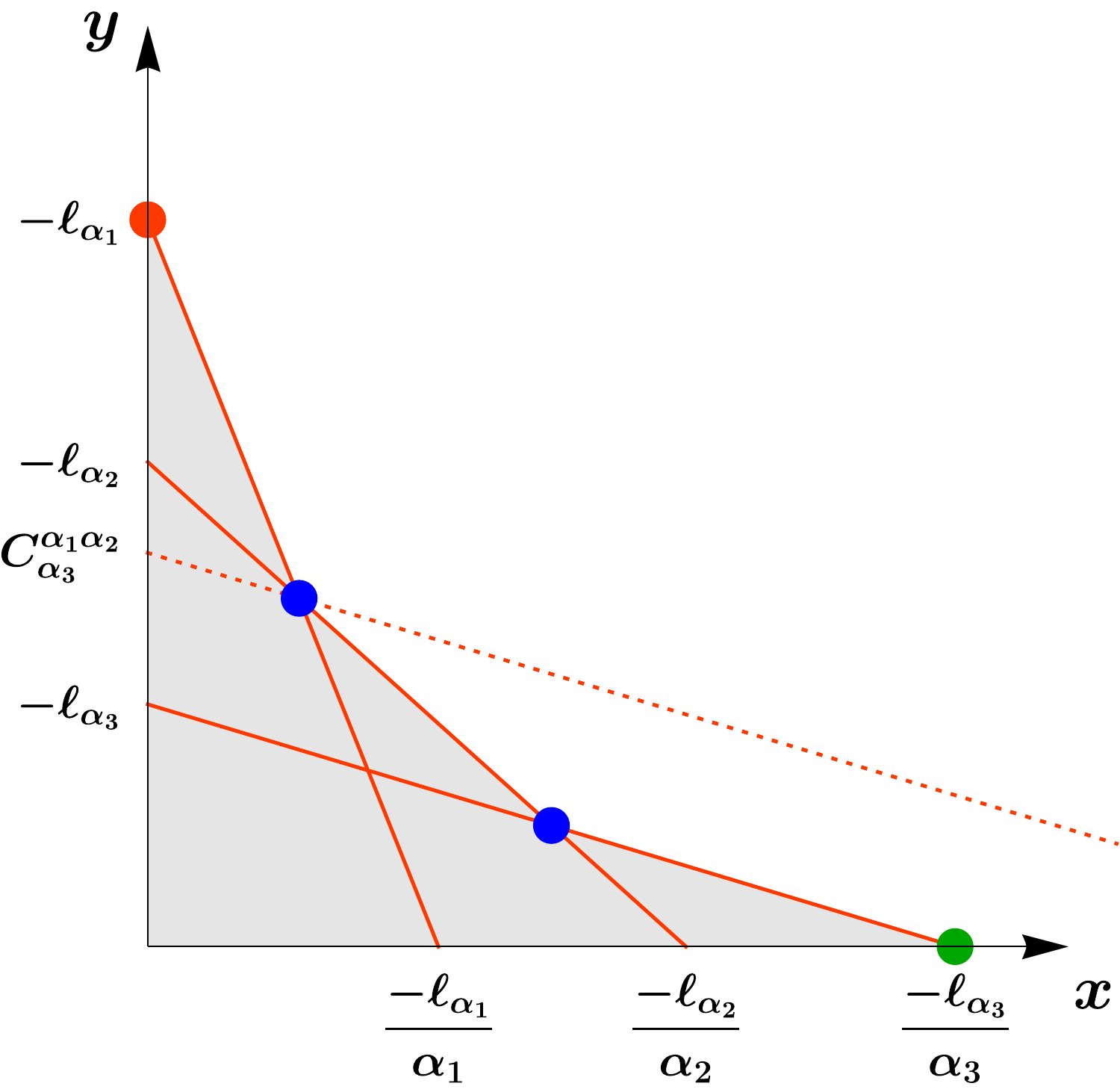}%
\caption{The Lund plane describing the region of phase space for the measurement of three angularities in which the most logarithms may be resummed. The various modes are denoted by the green (collinear), orange (soft) and blue (collinear-soft) dots. The dotted line serves to indicate the point $C_{\al_3}^{\al_1 \al_2}$, which shows up in the boundaries of the region of phase space.}
\label{fig:LundPlane3}
\end{figure}

This method of finding all relevant regions of phase space can be generalized to the simultaneous measurement of an arbitrary number of angularities. There is only one additional subtlety that has to be taken into account when more than two angularities are considered, which we illustrate in \fig{LundPlane3} for three angularities with parameters $\al_1 > \al_2 > \al_3$. If the line corresponding to $\ell_{\al_3}$ were to be placed above the position indicated by the dotted line, the angularity $\ell_{\al_2}$ would no longer be connected to the boundary of the region in which emissions are forbidden and hence not affect the cross section. The point at which the dotted line crosses the $y$-axis is given by
%%%
\begin{align}
C^{\al_1 \al_2}_{\al_3} = \frac{1}{\al_1-\al_2}\Bigl[(\al_2-\al_3)\ell_{\al_1} - (\al_1 - \al_3)\ell_{\al_2}\Bigr]\,.
\end{align}
%%%

The phase space of a cross section involving an arbitrary number of angularities $n$ can be divided into various regimes, as listed explicitly in \eq{regimes_2_angularities} for $n=2$. The regime in which the largest number of independent logarithms occur, is the one for which the edge of the forbidden (gray) region in the Lund plane involves every line corresponding to an individual angularity. For $n=2$, this corresponds to the center panel in \fig{LundPlane2} and for $n=3$, this situation is depicted in \fig{LundPlane3}. This region will be denoted by $R_n(\al_1,\ldots,\al_n)$, and its boundaries in phase space are given by
%%%
\begin{alignat}{2}
&y\text{-conditions:} \quad &-\ell_{\alpha_1} &> - \ell_{\alpha_2}\,, \quad C^{\alpha_1\alpha_2}_{\alpha_3} > -\ell_{\alpha_3}\,, \quad \ldots\,, \quad C^{\alpha_{n-2}\alpha_{n-1}}_{\alpha_n} > -\ell_{\alpha_n}\,, \nn\\
&x\text{-conditions:} \quad &- \frac{\ell_{\alpha_n}}{\alpha_{n}} &> -\frac{\ell_{\alpha_{n-1}}}{\alpha_{n-1}}\,, \quad \ldots\,, \quad - \frac{\ell_{\alpha_2}}{\alpha_2} > - \frac{\ell_{\alpha_1}}{\alpha_1}\,.
\end{alignat}
%%%
The first line consists of the $n-1$ conditions on the hierarchy between the points at which each line in the Lund plane crosses the $y$-axis. The second line contains the $n-1$ conditions on the hierarchy between the points at which the lines cross the $x$-axis. As the conditions consist solely of inequalities, this region in phase space is $n$-dimensional and will be called the ``bulk''.

Regions that involve fewer logarithms can be obtained by raising or lowering the point where an angularity crosses either axis in the Lund plane, such that two modes (the colored dots) overlap. This can be seen explicitly in \fig{LundPlane2} by starting from the center panel and raising $-\ell_{\al_2}$ until it reaches $-\ell_{\al_2} = -\ell_{\al_1}$ in the right panel, sliding the mode indicated by the blue dot up to the orange dot in the process. In full generality, the boundaries of a region in phase space involving a subset of angularities $\ell_{\bt_1},\ldots,\ell_{\bt_m}$ with $m < n$ and $\{\beta_1,\ldots,\beta_m\} \subset \{\al_1,\ldots,\al_n\}$ are found to be\footnote{Note that subsequent ${\bt_i}$ and ${\bt_{i+1}}$ do not necessarily correspond to consecutive $\al_j$, although we still adhere to the convention $\bt_1 > \ldots >\bt_m$.}
%%%
\begin{alignat}{3} \label{eq:boundary_conditions_n_angularities}
&y\text{-conditions:}  &-\ell_{\beta_1} &> - \ell_{\beta_2}\,, \quad C^{\beta_1\beta_2}_{\beta_3} > -\ell_{\beta_3}\,, \quad \ldots\,, \quad C^{\beta_{m-2}\beta_{m-1}}_{\beta_m} > -\ell_{\beta_m}\,, \nn\\
&x\text{-conditions:} &- \frac{\ell_{\beta_m}}{\beta_{m}} &> -\frac{\ell_{\beta_{m-1}}}{\beta_{m-1}}\,, \quad \ldots\,, \quad - \frac{\ell_{\beta_2}}{\beta_2} > - \frac{\ell_{\beta_1}}{\beta_1}\,, \nn\\
&B\text{-conditions:} \quad  &-\ell_{\beta_1} &= - \ell_{\alpha_i}  \qquad \text{for every } \alpha_i \text{ with }  \alpha_i > \beta_1\,, \nn\\
& &  C^{\beta_j \beta_{j+1}}_{\alpha_i} &= -\ell_{\alpha_i} \qquad \text{for every } \alpha_i \text{ with } \beta_j > \alpha_i > \beta_{j+1}\,, \nn\\
& & -\frac{\ell_{\beta_m}}{\beta_m} &= - \frac{\ell_{\alpha_i}}{\al_i} \qquad \text{for every } \alpha_i \text{ with }  \beta_m > \alpha_i\,,
\end{alignat}
%%%
where the $B$-conditions (boundary-conditions) contain all restrictions on the angularities that are only connected to the boundary of the shaded area in the Lund plane through a single point, i.e.~the angularities not involved in the region. As any such regime is characterized by $n-m$ equalities, it represents an $m$-dimensional region in the $n$-dimensional phase space, denoted by $R_m(\bt_1,\ldots,\bt_m)$. By considering all possible combinations of angularities it then follows that there are $\binom{n}{n-m}$ distinct regions of dimension $m$ in the phase space of $n$ angularities.

%===============================================================================
\subsection{Factorization formulas}
\label{subsec:fact}
%===============================================================================

The analytical resummation will be performed by making use of the Soft-Collinear Effective Theory (SCET)~\cite{Bauer:2000ew,Bauer:2000yr,Bauer:2001ct,Bauer:2001yt}, which describes the infrared limit of QCD. The relevant degrees of freedom are determined by the process and measurements under consideration. The version of SCET that involves the collinear and soft modes in \eq{modes_scaling_angularities}, known as SCET$_{\text{I}}$, correctly describes the regions of phase space dominated by a single angularity, e.g. the left and right panels in \fig{LundPlane2}. Regions of phase space involving multiple angularities (such as the middle panel in \fig{LundPlane2}) contain additional collinear-soft modes and are correctly described by SCET$_+$~\cite{Procura:2014cba,Bauer:2011uc,Larkoski:2015zka,Pietrulewicz:2016nwo}.

As the various modes in SCET are decoupled at the level of the Lagrangian~\cite{Bauer:2001yt}, cross sections may be factorized into products or convolutions of perturbative functions as long as the contributions of the various modes to the measurements can be shown to factorize as well. In general, each of these functions contains logarithms of the ratio of its inherent, natural scale and the common scale $\mu$. By solving their RGEs, they may be evaluated at their natural scales (where the logarithms are minimized) and then evolved towards a common scale $\mu$, resumming all the large logarithms in the process.

The relevant degrees of freedom for the bulk regime for $n=2$ angularities are shown in the middle panel of \fig{LundPlane2}. The orange dot represents the (ultra)soft mode, the green dot the collinear mode, and the blue dot corresponds to a collinear-soft mode~\cite{Bauer:2011uc,Procura:2014cba}, which contributes to the measurement of both angularities. For our process of interest, $e^+e^- \to \text{dijets}$, there are two distinct collinear directions corresponding to the two jets, and hence also two corresponding collinear and collinear-soft modes. The factorization formula for this regime was derived using SCET in \refscite{Procura:2014cba,Procura:2018zpn} and reads
%%%
\begin{align} \label{eq:factorization_bulk_two}
\frac{\df^2 \si^{R_2(\al_1,\al_2)}}{\df Q^{\al_1} e_{\al_1}\, \df Q^{\al_2}e_{\al_2}} &= 
H(Q^2,\mu) \,
S(Q^{\al_1} e_{\al_1},\mu) \underset{\al_1}{\otimes}
\cS(Q^{\al_1} e_{\al_1}, Q^{\al_2} e_{\al_2},\mu) \\
&\quad \underset{\al_1,\al_2}{\otimes} 
\cS(Q^{\al_1} e_{\al_1}, Q^{\al_2} e_{\al_2},\mu) 
\underset{\al_2}{\otimes} 
J(Q^{\al_2} e_{\al_2},\mu) \underset{\al_2}{\otimes} 
J(Q^{\al_2} e_{\al_2},\mu) \nn\\
& \equiv H(Q^2,\mu) \,
S(Q^{\al_1} e_{\al_1},\mu) \underset{\al_1}{\otimes}
\bigl[\cS(Q^{\al_1} e_{\al_1}, Q^{\al_2} e_{\al_2},\mu)\bigr]^2\! \underset{\al_2}{\otimes} 
\bigl[J(Q^{\al_2} e_{\al_2},\mu)\bigr]^2,\nn
\end{align}
where we have defined convolutions between two functions $f$ and $g$ through 
%%%
\begin{align}
f(Q^\al e_{\al},\ldots) \underset{\al}{\otimes} g(Q^\al e_{\al},\ldots) \equiv \int \! \df (Q^\al e_\al')\, f(Q^\al e_{\al} - Q^\al e_{\al}',\ldots)\, g(Q^\al e_{\al}',\ldots)\,.
\end{align}
%%%
Here the dots represent possible additional arguments. Furthermore, the short-hand notations 
%%%
\begin{align}
\bigl[J(Q^\al e_{\al},\mu)\bigr]^2 &\equiv J(Q^\al e_{\al},\mu) \underset{\al}{\otimes} J(Q^\al e_{\al},\mu)\, ,\nn\\
\bigl[\cS(Q^{\al_1}e_{\al_1},Q^{\al_2}e_{\al_2},\mu)\bigr]^2 &\equiv \cS(Q^{\al_1}e_{\al_1},Q^{\al_2}e_{\al_2},\mu) \underset{\al_1,\al_2}{\otimes} \cS(Q^{\al_1}e_{\al_1},Q^{\al_2}e_{\al_2},\mu)\,,
\end{align}
%%%
are employed, where $\underset{\al_1,\al_2}{\otimes}$ indicates a convolution in both $e_{\al_1}$ and $e_{\al_2}$. In \eq{factorization_bulk_two}, the hard function $H(Q^2,\mu)$ contains the Born cross section and virtual corrections to the hard scattering. The jet function $J(e_\al,\mu)$ describes collinear radiation, the soft function $S(e_\al,\mu)$ encodes the contribution from soft radiation, and $\cS(e_{\al_1},e_{\al_2},\mu)$ is the collinear-soft function. 

The region of phase space represented by the left panel of \fig{LundPlane2} is reached through raising $-\ell_{\al_1}/\al_1$ or lowering $-\ell_{\al_2}/\al_2$ until the two are equal, joining the collinear-soft mode with the collinear mode in the process. The factorization formula of this region then no longer contains any collinear-soft functions, but instead involves jet functions depending on both angularities
%%%
\begin{align}
\frac{\df^2 \si^{R_1(\al_1)}}{\df Q^{\al_1} e_{\al_1}\, \df Q^{\al_2} e_{\al_2}} = H(Q^2,\mu) \,
S(Q^{\al_1} e_{\al_1},\mu) &\underset{\al_1}{\otimes}
\bigl[J(Q^{\al_1} e_{\al_1},Q^{\al_2} e_{\al_2},\mu)\bigr]^2\,.
\end{align}
%%%
The consistency relation between the double-differential jet function and the convolution between the collinear-soft and single-differential jet function that this implies was verified explicitly at one-loop order in \refcite{Procura:2018zpn}. In this regime, soft radiation does not contribute to the measurement $e_{\al_2}$ and the factorization formula is simply a more differential version of the factorization formula for the sole measurement of $e_{\al_1}$.

Analogously, to obtain the factorization formula describing the region of phase space depicted in the right-most panel in \fig{LundPlane2}, the soft and collinear-soft functions merge into a more differential soft function to yield
%%%
\begin{align}
\frac{\df^2 \si^{R_1(\al_2)}}{\df Q^{\al_1} e_{\al_1}\, \df Q^{\al_2} e_{\al_2}} = H(Q^2,\mu) \,
S(Q^{\al_1} e_{\al_1},Q^{\al_2} e_{\al_2},\mu) &\underset{\al_2}{\otimes}
\bigl[J(Q^{\al_2} e_{\al_2},\mu)\bigr]^2\,.
\end{align}
%%%
This is a more differential version of the factorization theorem for the single-differential cross section in $e_{\al_2}$ and only resums large logarithms involving this angularity. 

The renormalization group equations of the perturbative functions occurring in the factorization formula in \eq{factorization_bulk_two} can be found in \app{resum}. Using the  expressions for the anomalous dimensions given in \eq{anomDims}, the consistency relation of the factorization formula in \eq{factorization_bulk_two}, given by
%%%
\begin{align}
0 &= \ga_H(Q^2,\mu)\de(Q^{\al_1}e_{\al_1})\de(Q^{\al_2}e_{\al_2}) + 2\ga_J(Q^{\al_2}e_{\al_2},\mu)\de(Q^{\al_1}e_{\al_1}) \nn\\
& \qquad  + 2 \ga_{\cS}(Q^{\al_1}e_{\al_1},Q^{\al_2}e_{\al_2},\mu) + \ga_S(Q^{\al_1}e_{\al_1},\mu)\de(Q^{\al_2}e_{\al_2})\,,
\end{align}
%%%
is indeed found to be satisfied.

For the measurement of $n$ angularities, the factorization formula for the cross section describing the bulk region $R_n(\al_1,\ldots,\al_n)$ follows from the modes appearing in the corresponding Lund plane. Specifically, there is a single soft mode, a single collinear mode (for each of the two collinear directions) and there are $n-1$ collinear-soft modes (per collinear direction), leading to the general factorization formula
%%%
\begin{align} \label{eq:factorization_n_angularities}
\frac{\df^n \si^{R_n(\al_1,\ldots,\al_n)}}{\df Q^{\al_1} e_{\al_1} \ldots \df Q^{\al_n} e_{\al_n}} &= 
H(Q^2,\mu) \,
S(Q^{\al_1} e_{\al_1},\mu) \underset{\al_1}{\otimes}
\bigl[\cS(Q^{\al_1} e_{\al_1}, Q^{\al_2} e_{\al_2},\mu)\bigr]^2 \underset{\al_2}{\otimes} \ldots \\
& \qquad \ldots \underset{\al_i}{\otimes} \bigl[\cS(Q^{\al_i} e_{\al_i}, Q^{\al_{i+1}} e_{\al_{i+1}},\mu)\bigr]^2 \underset{\al_{i+1}}{\otimes} \ldots \nn\\
& \qquad \ldots \underset{\al_{n-1}}{\otimes} \bigl[\cS(Q^{\al_{n-1}} e_{\al_{n-1}}, Q^{\al_n} e_{\al_n},\mu)\bigr]^2
\underset{\al_n}{\otimes}
\big[J(Q^{\al_n} e_{\al_n},\mu)\big]^2\,. \nn
\end{align}
%%%
When taking derivatives of this expression with respect to $\mu$, the anomalous dimensions of all intermediate collinear-soft functions effectively combine into a single collinear-soft anomalous dimension involving the angularities $e_{\alpha_1}$ and $e_{\alpha_n}$, leading to the conclusion that this factorization formula also obeys the corresponding consistency relation. Factorization formulas corresponding to regions that involve fewer angularities are again obtained by merging two degrees of freedom, i.e.~merging two functions into a single, more differential function. These can involve more than two angularities, but are not required to obtain cross sections at NLL accuracy, since only tree-level expressions are needed for the functions and the anomalous dimensions smoothly merge. Specifically, the tree-level expression of each function is simply a product of delta functions of its arguments, and the more differential functions that arise due to the merging of modes thus do not give rise to any different results for the cross section. This then implies that the cross section for a specific region of interest does not depend on the total set of angularities that are measured, but instead only on the subset of angularities that occur in said region. 

%===============================================================================
\subsection{Resummation}
\label{subsec:resum}
%===============================================================================

The large logarithms in a factorization formula, such as \eq{factorization_n_angularities}, can be resummed by evaluating each ingredient at its natural scale (where its logarithms are minimized) and then evolving them to a common scale. To solve the RGEs in \eq{RGEs} that involve a convolution, it is convenient to switch to a conjugate space. Picking Laplace space, the transformation of a function $f(t)$ is defined through
%%%
\begin{align}
\tilde{f}(s) \equiv \text{LT}[f(t)] \equiv \int_0^\infty \! \df t\, e^{-st} f(t)\,,
\end{align}
%%%
where LT$[\ldots]$ denotes the Laplace transform and $s$ is the variable conjugate to $t$. The plus distributions that appear are defined as
%%%
\begin{align}
\cL_n(x) \equiv \biggl[\frac{\theta(x)\ln^n(x)}{x}\biggr]_+ \qquad \text{and} \qquad \cL^a(x) \equiv \biggl[\frac{\theta(x)}{x^{1-a}}\biggr]_+\,.
\end{align}
%%%
The transformations of these distributions that are required up to NLL, are given by
%%%
\begin{align}
\text{LT}[\de(t)] &= 1\,, \qquad \text{LT}[\cL_0(t)] = -\ln(s\, e^{\ga_E}) \qquad \text{and} \qquad \text{LT}[\cL^a(t)] = - \frac{1}{a} + \frac{\Gamma(a)}{s^a}\,.
\end{align}
%%%
Solving the RGEs in Laplace space, inserting these results in the cross section in \eq{factorization_n_angularities} (at NLL) and transforming back to momentum space then yields the resummed cross section. To ensure that our procedure allows the recovery of the inclusive cross section upon integration over all the angularities that the differential cross section depends on, we perform the resummation at the level of the cumulative cross section~\cite{Abbate:2010xh, Almeida:2014uva, Bertolini:2017eui}. To obtain the cumulative cross section, we integrate over each angularity $e_{\al}$ up to some cut-off $e_{\al}^\cut$, which is the basis for our numerical implementation. Leaving this ``cut" superscript implicit, the cumulative cross section is 
%%%
\begin{align} \label{eq:resummed_cumulative}
\sigma_{R_n(\al_1,\ldots,\al_n)} &= \hat{\si}_0\frac{\exp\bigl[K_H + K_S^{0}(\al_1) + 2 K_J -\ga_E(2\eta_J + \eta_S^{n-1}(\al_n) )\bigr]}{\Ga[1 + 2\eta_J + \eta_S^{n-1}(\al_n)]} \nn\\
&\quad \times \Bigl(\frac{Q}{\mu_H}\Bigr)^{\eta_H}\Bigl(\frac{Qe_{\al_1}}{\mu_S}\Bigr)^{\eta_S^{0}(\al_1)}\Bigl(\frac{Qe_{\al_n}^{1/\al_n}}{\mu_{J}}\Bigr)^{2\al_n\eta_J}  \nn\\
& \quad \times \prod_{i=1}^{n-1} \Biggl\{\frac{\exp\bigl[K_S^{i}(\al_{i+1}) - K_S^{i}(\al_i) - \ga_E\,  \eta_S^{i-1,i}(\al_i)\bigr]}{\Ga[1 +  \eta_S^{i-1,i}(\al_i)]} \nn\\
& \quad \times \biggl(e_{\al_i}^{\al_{i+1}-1}e_{\al_{i+1}}^{1-\al_i}\Bigl(\frac{Q}{\mu_{i,i+1}}\Bigr)^{\al_{i+1} - \al_i}\biggr)^{\frac{2\eta_\Ga(\mu_{i,i+1},\mu)}{(\al_i - 1)(1 - \al_{i+1})}}\Biggr\}\,.
\end{align}
%%%
Here we have defined 
%%%
\begin{alignat}{2}
K_H &\equiv -4K_\Ga(\mu_H,\mu) + K_{\ga_H}(\mu_H,\mu)\,,& \qquad \eta_H &\equiv 4\eta_\Ga(\mu_H,\mu)\,, \nn\\
K_J &\equiv \frac{2\al_n}{\al_n-1}K_\Ga(\mu_J,\mu) + K_{\ga_J}(\mu_J,\mu)\,,& \qquad \eta_J &\equiv \frac{2}{1-\al_n}\eta_\Ga(\mu_J,\mu)\,, \nn\\
K_S^{i}(\al_j) &\equiv -\frac{4}{\al_j - 1}K_\Ga(\mu_{i,i+1},\mu)\,,& \qquad \eta_S^{i}(\al_j) &\equiv \frac{4}{\al_j - 1}\eta_\Ga(\mu_{i,i+1},\mu)\,,
\end{alignat}
%%%
in terms of the evolution kernels that can be found in \app{resum}. Furthermore, the notation 
%%%
\begin{align}
\eta_S^{i-1}(\al_i) - \eta_S^{i}(\al_i) = \frac{4}{\al_i - 1}\int_{\al_s(\mu_{i-1,i})}^{\al_s(\mu_{i,i+1})} \! \frac{\df \al_s'}{\beta(\al_s')}\, \Ga_{\rm cusp}(\al_s') \equiv \eta_S^{i-1,i}(\al_i)\,,
\end{align}
%%%
has been employed to simplify the result. The natural scales of the functions at which the large logarithms are minimized depend on the (sub)set of angularities under consideration. Denoting this set of angularities by $\bt_1,\ldots,\bt_m$, the various natural scales are given by
%%%
\begin{alignat}{2}
\mu_H &= Q\,, \qquad &\mu_S &\equiv \mu_{0,1} = Qe_{\beta_1}\,, \nn\\ 
\mu_J &= Q e_{\beta_m}^{1/\beta_m}\,, \qquad &\mu_\cS(\beta_i,\beta_{i+1}) &\equiv \mu_{i,{i+1}} = Q \biggl(\frac{e_{\beta_{i}}^{1-\beta_{i+1}}}{e_{\beta_{i+1}}^{1-\beta_{i}}}\biggr)^{\frac{1}{\beta_{i} - \beta_{i+1}}},
\end{alignat}
%%%
where again we have suppressed the superscript ``cut" on the angularities.
To avoid the Landau pole in our numerical implementation, we freeze the value of $\al_s$ below 2 $\GeV$. 

%===============================================================================
\subsection{Power corrections}
\label{subsec:pc}
%===============================================================================

The power corrections to each factorization formula can be determined by considering the ratio of scales involved in the functions that are merged into a more differential function when transitioning towards a lower-dimensional region in phase space. The lower-dimensional region will be referred to as a `daughter region' with respect to the higher-dimensional `parent region'. The measurement of three angularities will be used as an example to display this procedure. The various regions of phase space and their $B$-conditions, i.e.~the equalities in \eq{boundary_conditions_n_angularities}, can be found in \tab{regions} for $\alpha_1 > \alpha_2 > \alpha_3$.
\begin{table}
\centering
\begin{tabular}{ l l l }
   \hline \hline \\[-8pt]
   Region $R_n(\alpha_1,\ldots,\alpha_n)$ & \multicolumn{2}{l}{Boundary conditions $B_n(\alpha_1,\ldots,\alpha_n)$} \\[2pt]
   \hline \hline \\[-10pt]
   $R_3(\alpha_1,\alpha_2,\alpha_3)$ & \multicolumn{2}{l}{-} \\[2pt]
   \hline \\[-12pt]
   $R_2(\alpha_1,\alpha_2)$ & \multicolumn{2}{l}{$e_{\alpha_3} = e_{\alpha_2}^{\alpha_3/\alpha_2}$} \\[2pt]
   $R_2(\alpha_1,\alpha_3)$ & \multicolumn{2}{l}{$e_{\alpha_2} = (e_{\alpha_1}^{\alpha_3-\alpha_2}/e_{\alpha_3}^{\alpha_2-\alpha_1})^{1/(\alpha_3-\alpha_1)}$} \\[2pt]
   $R_2(\alpha_2,\alpha_3)$ & \multicolumn{2}{l}{$e_{\alpha_1} = e_{\alpha_2}$} \\[2pt]
   \hline \\[-12pt]
   $R_1(\alpha_1)$ & $e_{\alpha_2} = e_{\alpha_1}^{\alpha_2/\alpha_1}$ &and $\quad e_{\alpha_3} = e_{\alpha_1}^{\alpha_3/\alpha_1}$ \\[2pt]
   $R_1(\alpha_2)$ & $e_{\alpha_1} = e_{\alpha_2}$ &and $\quad e_{\alpha_3} = e_{\alpha_2}^{\alpha_3/\alpha_2}$\\[2pt]
   $R_1(\alpha_3)$ & $e_{\alpha_1} = e_{\alpha_3}$ &and $\quad e_{\alpha_2} = e_{\alpha_3}$ \\[2pt]
   \hline \hline
\end{tabular}
\caption{The boundary conditions of the various regions in the three-angularity phase space with $\alpha_1 > \alpha_2 > \alpha_3$.}
\label{tab:regions}
\end{table}%
In general, we denote the power corrections from an $n$-dimensional parent region $R_n(\al_1,\ldots,\al_n)$ towards an $(n-1)$-dimensional daughter region $R_{n-1}(\al_1,\ldots,\al_i,\al_{i+1},\ldots,\al_n)$ by $P_n(\alpha_1,\ldots,\alpha_n; \alpha_i)$, where the argument after the semicolon indicates the angularity that the daughter region lacks with respect to the parent region. Using this notation, the power corrections from the one-dimensional regions towards the fixed-order region are given by~\cite{Procura:2018zpn}
%%% 
\begin{align}
P_1(\alpha_i; \alpha_i) &= e_{\alpha_i}^{\min[2/\alpha_i,1]}\,.
\end{align}
%%%
The power corrections of the factorization formula of a two-dimensional region towards the one-dimensional daughter regions are given by 
%%%
\begin{align}
P_2(\alpha_i,\alpha_j; \alpha_j) &= \Bigl(\frac{\mu_\cS(\alpha_i,\alpha_j)}{\mu_J(\alpha_i)}\Bigr)^\#\,, \nn\\
P_2(\alpha_i,\alpha_j; \alpha_i) &= \Bigl(\frac{\mu_\cS(\alpha_i,\alpha_j)}{\mu_S(\alpha_j)}\Bigr)^\#\,.
\end{align}
%%%
The powers denoted by $\#$ are (different) constants, which may be fixed by demanding that the power corrections should reduce to those of the one-dimensional region at the corresponding boundary, i.e.
%%%
\begin{align}
\bigl[P_2(\alpha_i,\alpha_j; \alpha_j)\bigr]_{B_1(\alpha_j)} &\stackrel{!}{=} P_1(\alpha_j; \alpha_j)\,, \nn\\
\bigl[P_2(\alpha_i,\alpha_j;\, \alpha_i)\bigr]_{B_1(\alpha_i)} &\stackrel{!}{=} P_1(\alpha_i; \alpha_i)\,,
\end{align}
%%%
where the boundary conditions $B_1$ can be found in \tab{regions}. Plugging in the various scales then yields the complete set of power corrections of the two-dimensional region
%%%
\begin{align}
P_2(\alpha_i,\alpha_j) &= \biggl\{ \biggl(\frac{e_{\al_j}}{e_{\al_i}^{\al_j/\al_i}}\biggr)^{\frac{\al_i}{\al_i-\al_j}\min[2/\al_j,1]} , \biggl(\frac{e_{\al_i}}{e_{\al_j}}\biggr)^{\frac{\al_i}{\al_i-\al_j}\min[2/\al_i,1]} \biggr\}\,.
\end{align}
%%%
The power corrections from the three-dimensional region towards any of the neighboring two-dimensional regions can be found in an analogous way. By again demanding that these power corrections should reduce to those of any other boundary theory, we obtain the set of power corrections of the three-dimensional region
%%%
\begin{align} \label{eq:PR_3}
P_3(\alpha_i,\alpha_j,\alpha_k) = \biggl\{ &\biggl(\frac{e_{\al_k}}{e_{\al_j}^{\al_k/\al_j}}\biggr)^{\frac{\al_j}{\al_j-\al_k}\min[2/\al_k,1]} , 
\biggl(\frac{e_{\al_i}^{\al_j - \al_k}e_{\al_k}^{\al_i-\al_j}}{e_{\al_j}^{\al_i-\al_k}}\biggr)^{\frac{\al_j}{(\al_j - \al_i)(\al_j - \al_k)}\min[2/\al_j,1]} , \nn\\
&\biggl(\frac{e_{\al_i}}{e_{\al_j}}\biggr)^{\frac{\al_i}{\al_i-\al_j}\min[2/\al_i,1]} \biggr\}\,.
\end{align}
%%%
This procedure is easily generalized to the case of $n$ angularities. There are three different types of power corrections that need to be considered, all of them already present for $n=3$. They are given by
%%%
\begin{align} \label{eq:PR_n}
P_n(\alpha_1,\ldots,\alpha_n; \alpha_n) &= \biggl(\frac{e_{\alpha_n}}{e_{\alpha_{n-1}}^{\alpha_n/\alpha_{n-1}}}\biggr)^{E_n(\alpha_1,\ldots,\alpha_n; \alpha_n)}, \nn\\
P_n(\alpha_1,\ldots,\alpha_n; \alpha_i) &= \biggl(\frac{e_{\alpha_{i-1}}^{\alpha_i-\alpha_{i+1}}e_{\alpha_{i+1}}^{\alpha_{i-1} - \alpha_i}}{e_{\alpha_{i}}^{\alpha_{i-1} - \alpha_{i+1}}}\biggr)^{E_n(\alpha_1,\ldots,\alpha_n; \alpha_{i})} \quad \text{for} \quad 2 \leq i \leq n-1\,, \nn\\
P_n(\alpha_1,\ldots,\alpha_n; \alpha_1) &= \biggl(\frac{e_{\alpha_1}}{e_{\alpha_{2}}}\biggr)^{E_n(\alpha_1,\ldots,\alpha_n; \alpha_1)}.
\end{align}
%%%
The powers $E$ can then be found through demanding
%%%
\begin{align}
\bigl[P_n(\alpha_1,\ldots,\alpha_n; \alpha_i)\bigr]_{B_{n-1}(\alpha_2,\ldots,\alpha_n)} &\stackrel{!}{=} P_{n-1}(\alpha_2,\ldots,\alpha_n; \alpha_i)\,, \nn\\
\bigl[P_n(\alpha_1,\ldots,\alpha_n; \alpha_1)\bigr]_{B_{n-1}(\alpha_1,\ldots,\alpha_{n-1})} &\stackrel{!}{=} P_{n-1}(\alpha_1,\ldots,\alpha_{n-1}; \alpha_1)\,,
\end{align}
%%%
for $2 \leq i \leq n$.

%===============================================================================
\subsection{Matching phase space regions}
\label{subsec:match}
%===============================================================================

The different regions of phase space that are found using Lund diagrams are each described by different cumulative cross sections. In order to obtain a combined prediction valid throughout phase space, these regions need to be matched to each another. For any given point in $n$-dimensional phase space, the combined cumulative cross section is defined as a linear combination of all possible regions that occur in phase space as
%%%
\begin{align} \label{eq:combined_cumulant}
\sigma(e_{\alpha_1},\ldots,e_{\alpha_n}) = \sum_{R_m} a_m(\beta_1,\ldots,\beta_m)\,\sigma^{R_m(\beta_1,\ldots,\beta_m)}\,,
\end{align}
%%%
where again $\beta_1,\ldots,\beta_m$ is any subset of the full set of angularities $\al_1,\ldots,\al_n$ and the dependence of the transition variables and the cross sections on the full set of angularities has been suppressed. The sum runs over all possible regions with $n \geq m \geq 0$ and the set of coefficients is normalized as
%%%
\begin{align}
\sum_{R_m} a_m(\beta_1,\ldots,\beta_m) = 1\,,
\end{align}
%%%
at every point in phase space spanned by the $n$ angularities under consideration. In principle, this includes the matching to the fixed-order region, denoted by $R_0$, but this matching is not performed here for simplicity, so that we simply set $a_0 = 0$. Following the approach in ref.~\cite{Echevarria:2018qyi}, the specific admixture of transition variables $a_i$ is determined by the size of the power corrections to the factorization formula in each region. We start by defining a transition function that smoothly interpolates between 0 and 1 as
%%%
\begin{align} \label{eq:f_Trans}
f_{\text{trans}}(x_i,x_f,x) &= 
\begin{cases}
0 \qquad &\text{if}\ x > x_i\,, \\
\sum_{j=0}^5 c_j(x_i,x_f)x^j/(x_f-x_i)^5 \qquad &\text{if}\ x_i \geq x > x_f\,, \\
1 \qquad &\text{if}\ x_f \geq x\,,
\end{cases}
\end{align}
%%%
where the constants $c_j$ are determined by demanding the continuity of $f_{\text{trans}}$ and its first and second derivative at both transition points $x_i$ and $x_f$. The explicit expressions obtained in this way are given by
%%%
\begin{alignat}{2}
c_0(x_i,x_f) &= -(10 x_f^2 x_i^3-5 x_f x_i^4+x_i^5)\,,& \qquad c_3(x_i,x_f) &= 10 (x_f^2+4 x_f x_i+x_i^2)\,, \nn\\
c_1(x_i,x_f) &= 30 x_f^2 x_i^2\,,& \qquad c_4(x_i,x_f) &= -15 (x_f+x_i)\,,  \nn\\
c_2(x_i,x_f) &= -30 (x_f^2 x_i+x_f x_i^2)\,,& \qquad c_5(x_i,x_f) &= 6\,.
\end{alignat}
%%%
The explicit values of the transition variables $a_m(\beta_1,\ldots,\beta_m)$ at a given point $p$ in the $n$-dimensional space spanned by $\ell_{\al_1},\ldots,\ell_{\al_n}$ are determined iteratively. All transition variables are initialized at 0. The region in which $p$ lies is determined through the conditions given in \eq{boundary_conditions_n_angularities}. If it lies outside of all regions, the transition variables are kept fixed at zero. If $p$ lies inside a certain region $R_m(\beta_1,\ldots,\beta_m)$ involving $m$ angularities, the following procedure is followed:
\begin{itemize}
\item The set of daughter regions involving $m-1$ angularities, obtained by removing any single angularity from $R_m(\beta_1,\ldots,\beta_m)$, is determined.
\item The shortest Euclidean distance in the space spanned by $\ell_{\alpha_1},\ldots,\ell_{\alpha_n}$ from the point $p$ towards each daughter region is determined using the method of Lagrange multipliers~\cite{Lagrange:1788}. These distances are translated to a number between $0$ and $1$ through \eq{f_Trans}, where the initial and final points $x_i$ and $x_f$ correspond to the distances where the power corrections are 10\% and 50\% respectively. The result of this procedure is denoted by $\tilde{a}_m(\beta_1,\ldots,\beta_m;\beta_j)$, where the angularity after the semicolon again indicates the angularity that is involved in the parent region, but not in the daughter region. 
\item The coefficient of the region $R_m(\beta_1,\ldots,\beta_m)$ is defined through
%%%
\begin{align} \label{eq:coefficientWeightDetermination}
a_m(\beta_1,\ldots,\beta_m) = 1 - \max_j[\tilde{a}_m(\beta_1,\ldots,\beta_m;\beta_j)]\,,
\end{align}
%%%
and a preliminary weight $b_{m-1}(\gamma_1,\ldots,\gamma_{m-1}; \beta_i)$ is assigned to each of the $m$ daughter regions $R_{m-1}(\gamma_1,\ldots,\gamma_{m-1})$. Here the $\beta_i$ after the semicolon in this case indicates the angularity that should be added to the set  $\{\gamma_1,\ldots,\gamma_{m-1}\} \subset \{\beta_1,\ldots,\beta_m\}$ to obtain the full set of angularities $\{\beta_1,\ldots,\beta_m\}$ on which the parent region depends. The preliminary weights are given by
%%%
\begin{align}
b_{m-1}(\gamma_1,\ldots,\gamma_{m-1}; \beta_i) = \frac{\tilde{a}_m(\beta_1,\ldots,\beta_m;\beta_i)\,[1-a_m(\beta_1,\ldots,\beta_m)]}{\sum_j[\tilde{a}_m(\beta_1,\ldots,\beta_m; \beta_j)]}\,.
\end{align}
%%%
\item For each of the daughter regions $R_{m-1}$, the steps above are repeated in order to determine the transition variables $a_{m-1}$. The only notable difference is that the right-hand side of \eq{coefficientWeightDetermination} is to be multiplied by a factor $0 \leq \tilde x \leq 1$, given by the sum of all the preliminary weights that the region under consideration might have inherited from all of its parent regions. For a region $R_{m-1}(\gamma_1,\ldots,\gamma_{m-1})$, this factor is then given by
\begin{align}
\tilde{x} = \sum_{i} b_{m-1}(\gamma_1,\ldots,\gamma_{m-1}; \beta_i)\,.
\end{align}
\end{itemize}
This procedure is repeated until all regions from $R_m$ down to $R_2$ have been considered. The transition variables of the regions $R_1(\beta_i)$ are then given by the sum of the preliminary weights
\begin{align}
a_{1}(\beta_i) = \sum_j b_1(\beta_i;\beta_j)\,,
\end{align}
that they might have inherited from any of their parent regions. After all the coefficients have been determined, the cumulative distribution can be obtained through \eq{combined_cumulant}. In some cases in our numerical implementation, the cumulative distribution turns out to slightly decrease towards the fixed-order region due to the finite bin size. To ensure that this does not lead to negative spectra upon differentiation, any such bins are set equal to the average of their neighboring bins.

%%%%%%%%%%%%%%%%%%%%%%%%%%%%%%%%%%%%%%%%%%%%%%%%%%%%%%%%%%%%%%%%%%%%%%%%%%%%%%%%
\section{Results}
\label{sec:results}
%%%%%%%%%%%%%%%%%%%%%%%%%%%%%%%%%%%%%%%%%%%%%%%%%%%%%%%%%%%%%%%%%%%%%%%%%%%%%%%%

This section contains results obtained through the reweighing procedure described in \sec{reweigh}. By default we show results from \Herwig 7.1.4 for leading order $e^+e^- \rightarrow \text{dijets}$ (excluding bottom and top quark jets) at center-of-mass energy $Q=1$~TeV. The final-state parton shower is turned on, but the initial-state QED radiation and modeling of hadronization are switched off. The two jets are obtained via the exclusive $k_t$ algorithm~\cite{Catani:1991hj} with the winner-take-all recombination scheme~\cite{Salam:WTAUnpublished,Bertolini:2013iqa} using the \FastJet package~\cite{Cacciari:2011ma}. We consider the set of angularities with exponent $\alpha_i = 0.2\times s$ with $s = 1, 2, \dots, 15$, and use $k=4$-body phase space for reweighing. We also show our analytic predictions, as well as those obtained from \Pythia 8.240.

\begin{figure}[t]
\centering
\includegraphics[width=0.45\textwidth]{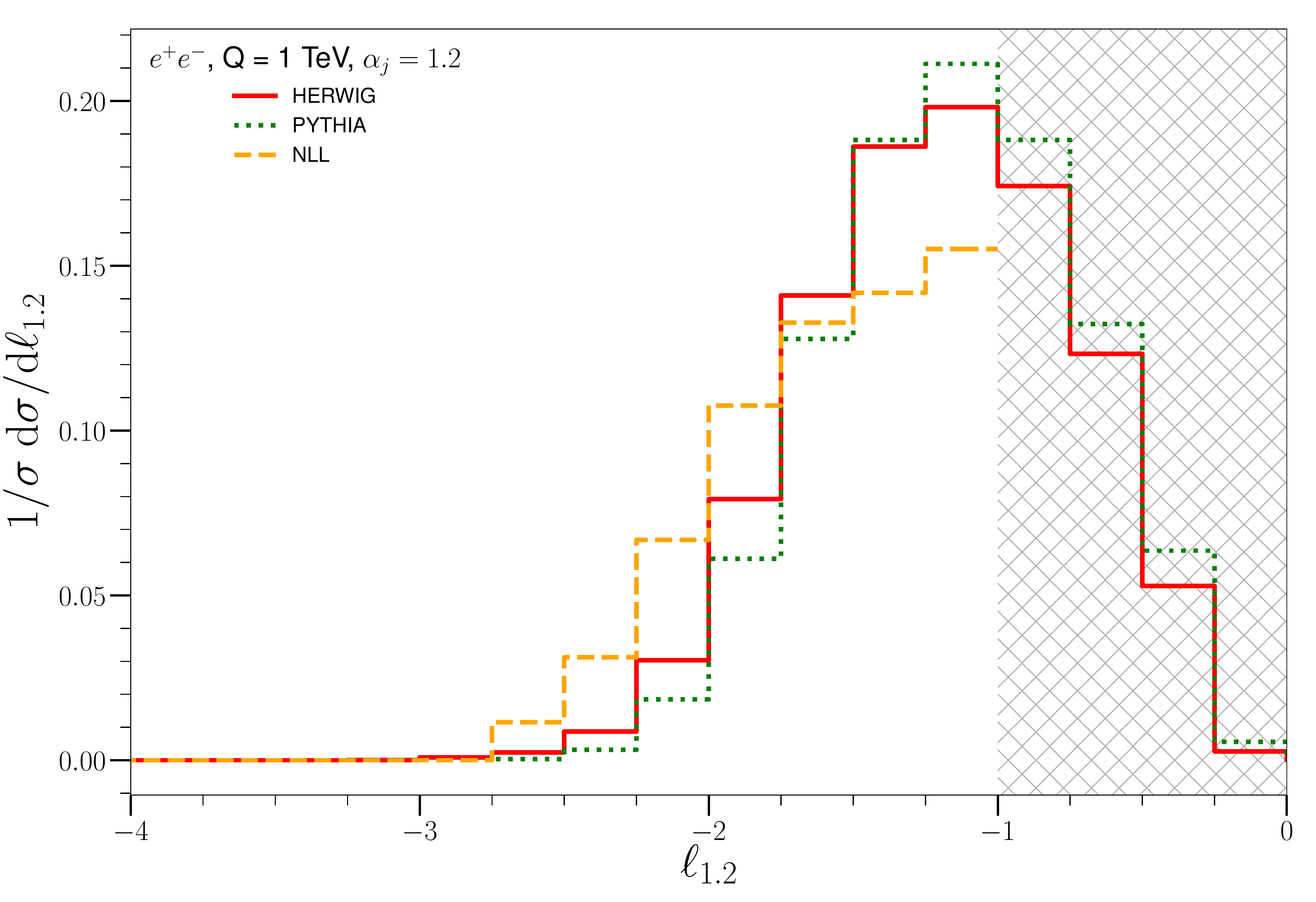}%
\includegraphics[width=0.45\textwidth]{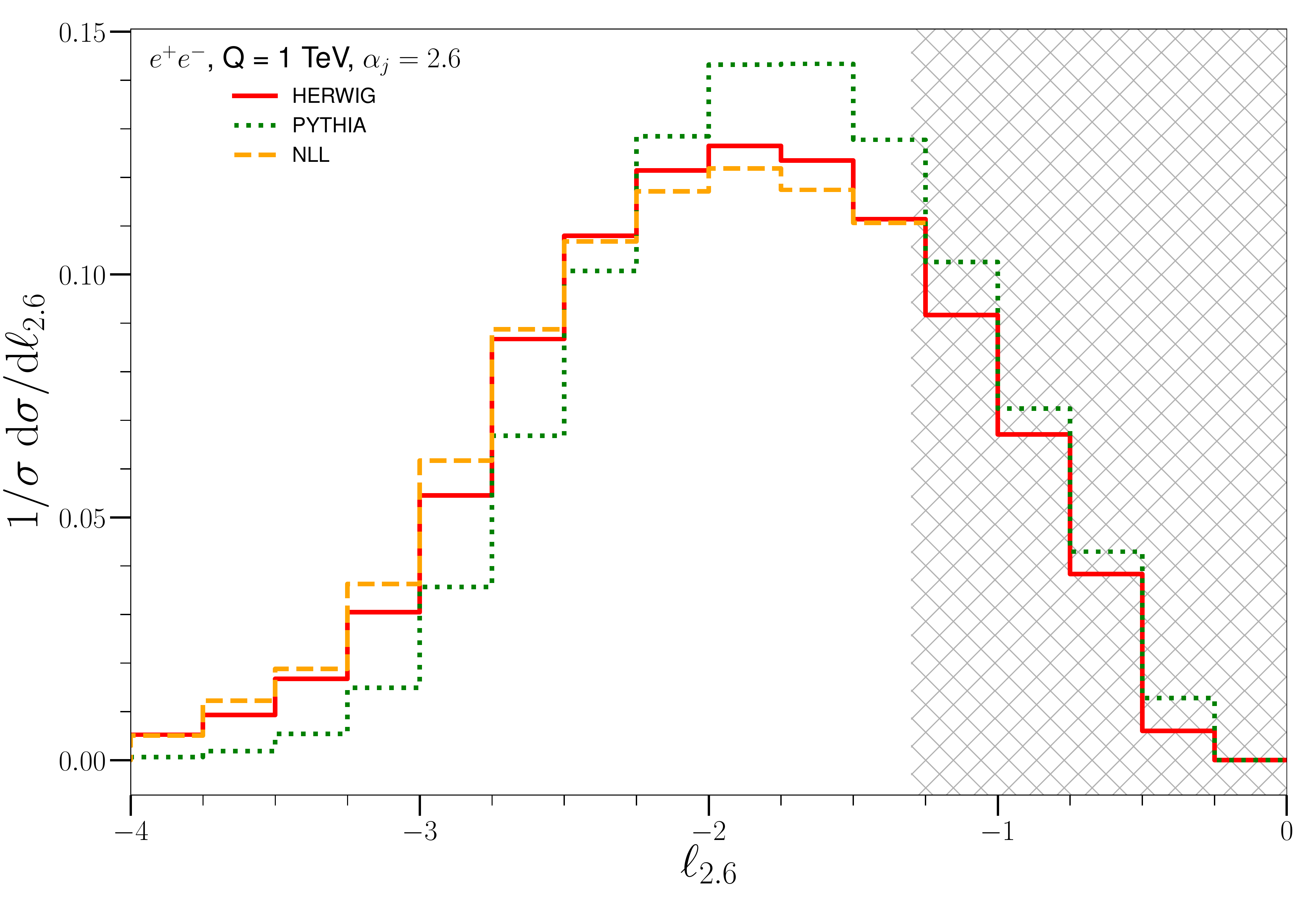}
\caption{The \Herwig (red), \Pythia (green dotted) and analytical NLL (yellow dashed) predictions for the $\ell_{1.2}$ and $\ell_{2.6}$ distributions. The fixed-order region has been grayed out and the analytical results have been normalised to the fraction of the area of the \Herwig results that lies to the left of that region.}
\label{fig:single1Dall}
\end{figure}

We begin by showing a comparison between the \Herwig, \Pythia and analytic predictions for the single angularity distribution in \fig{single1Dall}.
We find good agreement between the \Herwig and \Pythia results. The analytical result agrees very well with the numerical results for the angularity $\ell_{2.6}$, but shows some deviations for $\ell_{1.2}$. The reason for this is that the resummation region gets squeezed between the fixed-order region and the non-perturbative region\footnote{While we do not include hadronization, this region is sensitive to the unphysical shower cut off.}.

\begin{figure}[t]
\centering
\includegraphics[width=0.45\textwidth]{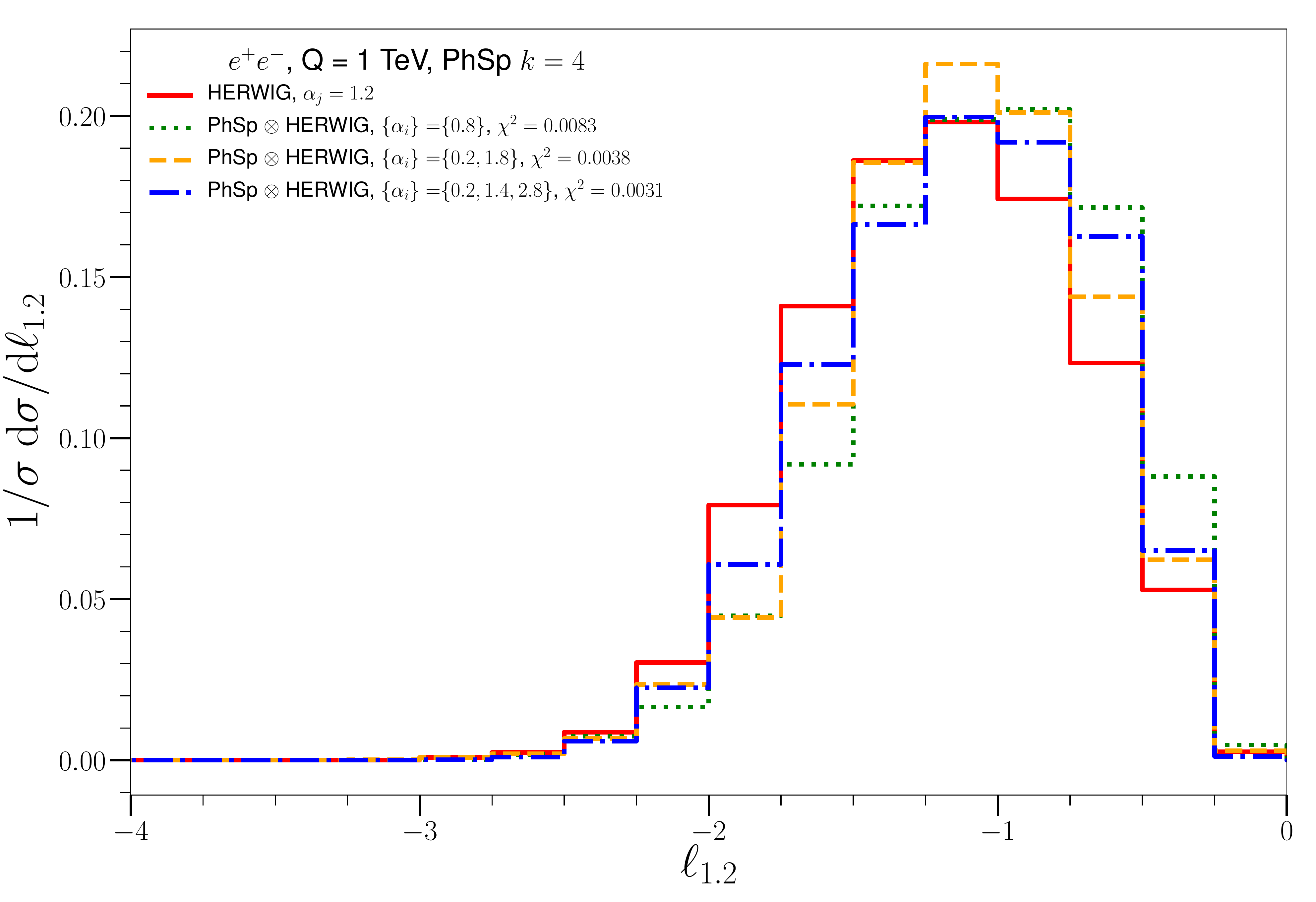}%
\includegraphics[width=0.45\textwidth]{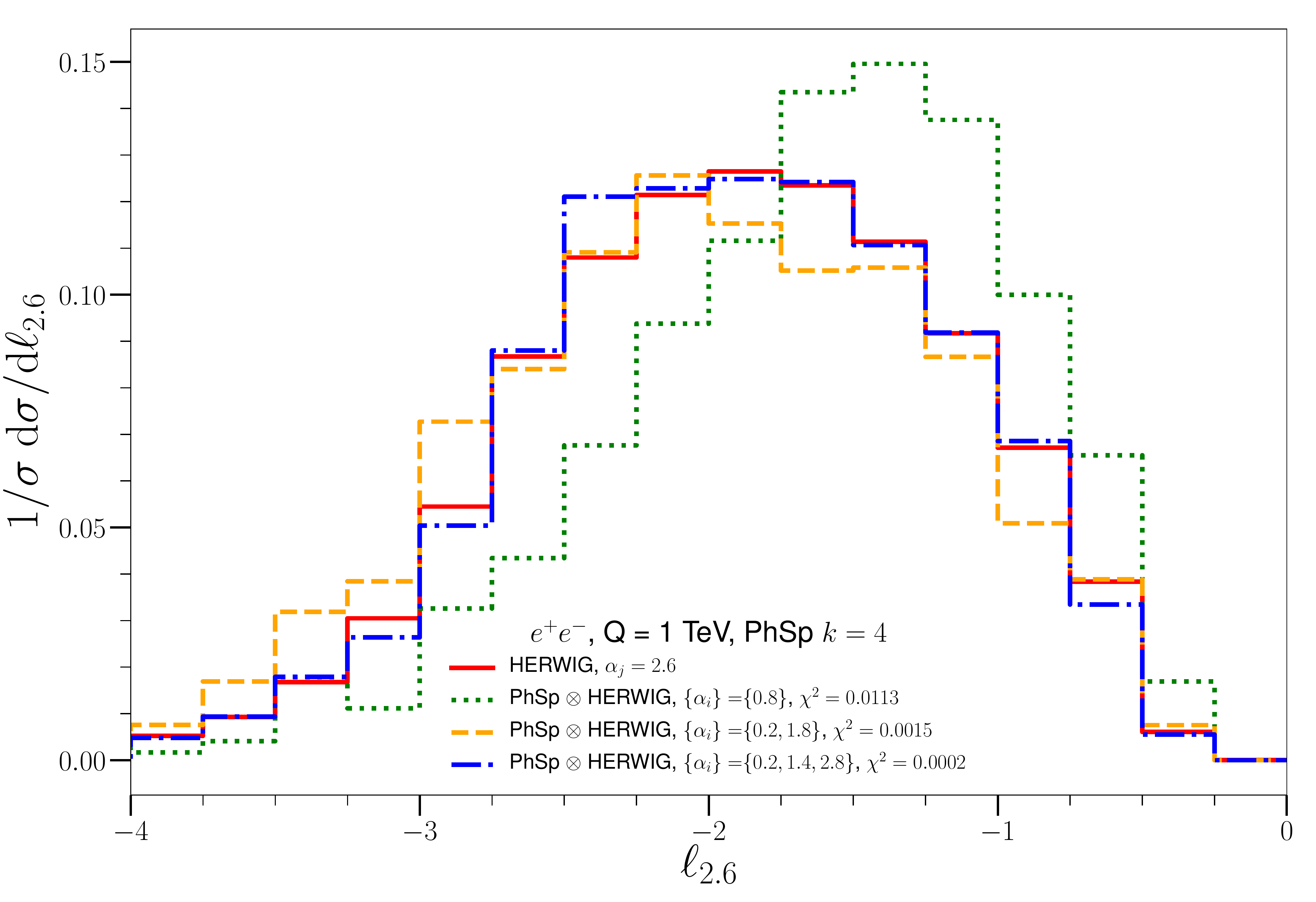}
\caption{Two examples of the reweighing procedure (with \Herwig) using the best $n=1$ (green dotted), $n=2$ (yellow dashed) and $n=3$ (blue dot-dashed) angularities yielding the global minimum indicated for each case. The red curve shows the distribution obtained directly from \Herwig 7. Left panel: The result for the $\alpha_j = 1.2$ exponent. In this case $n=1,2,3$ perform similarly. 
Right panel: The result for $\alpha_j = 2.6$. A clear improvement can be observed here as $n$ is increased.}
\label{fig:single}
\end{figure}

In \fig{single}, we show results for two examples, obtained through reweighing with the best possible set of $n=1,2,3$ angularities. In the left panel we show the results for $\alpha_j = 1.2$, where the performance for each $n$ is comparable. In the right panel, where we show $\alpha_j = 2.6$, there is a dramatic improvement going from the best possible reweighing with $n=1$ angularity to the best result for $n=2$ reweighed angularities. We stress that the best set of angularities is obtained through a global minimization and is thus not optimized for any specific  $\alpha_j$. The improvement from $n=2$ to $n=3$ is substantial, although not as dramatic. 

\begin{figure}[t]
\centering
\includegraphics[width=0.45\textwidth]{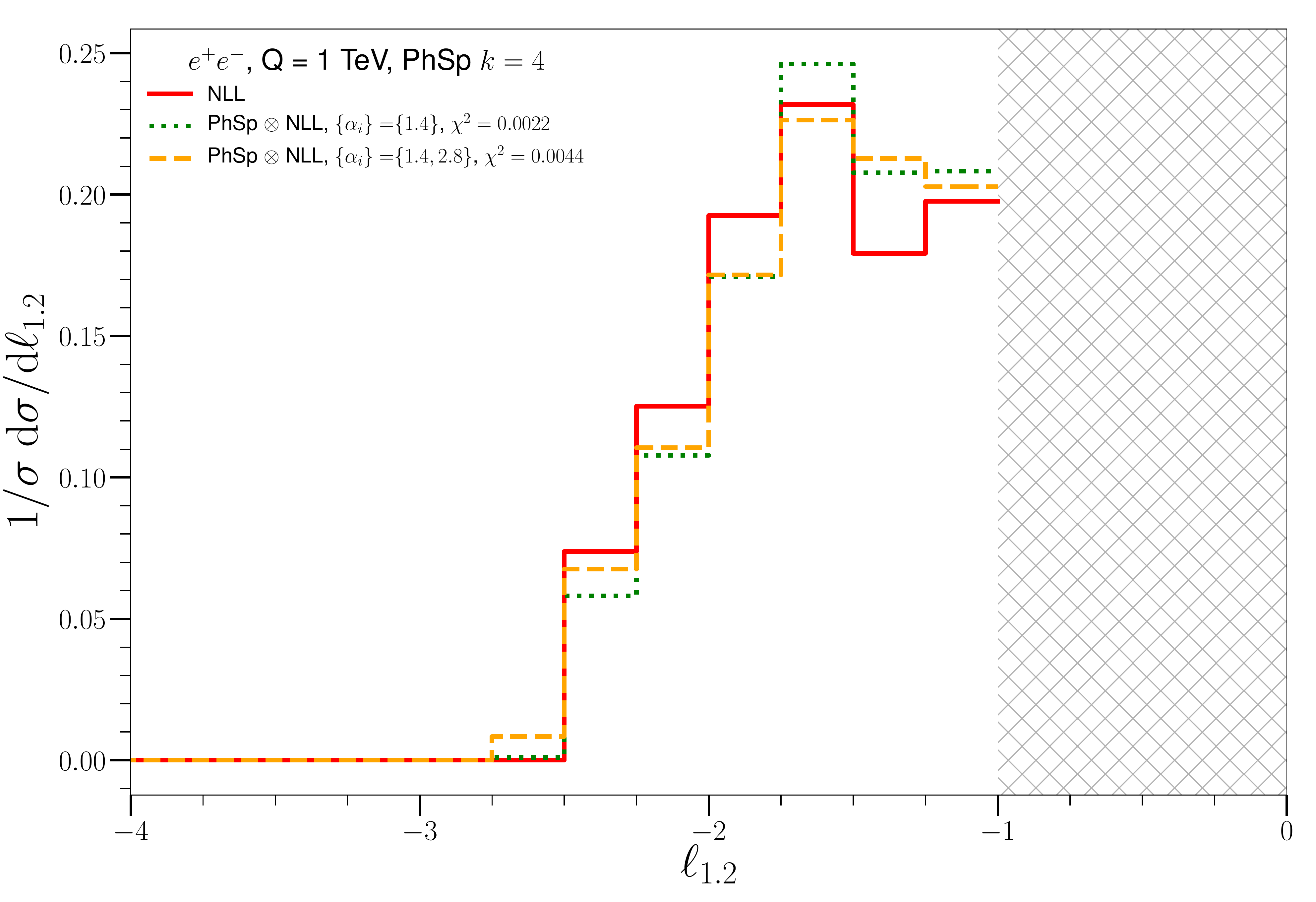}%
\includegraphics[width=0.45\textwidth]{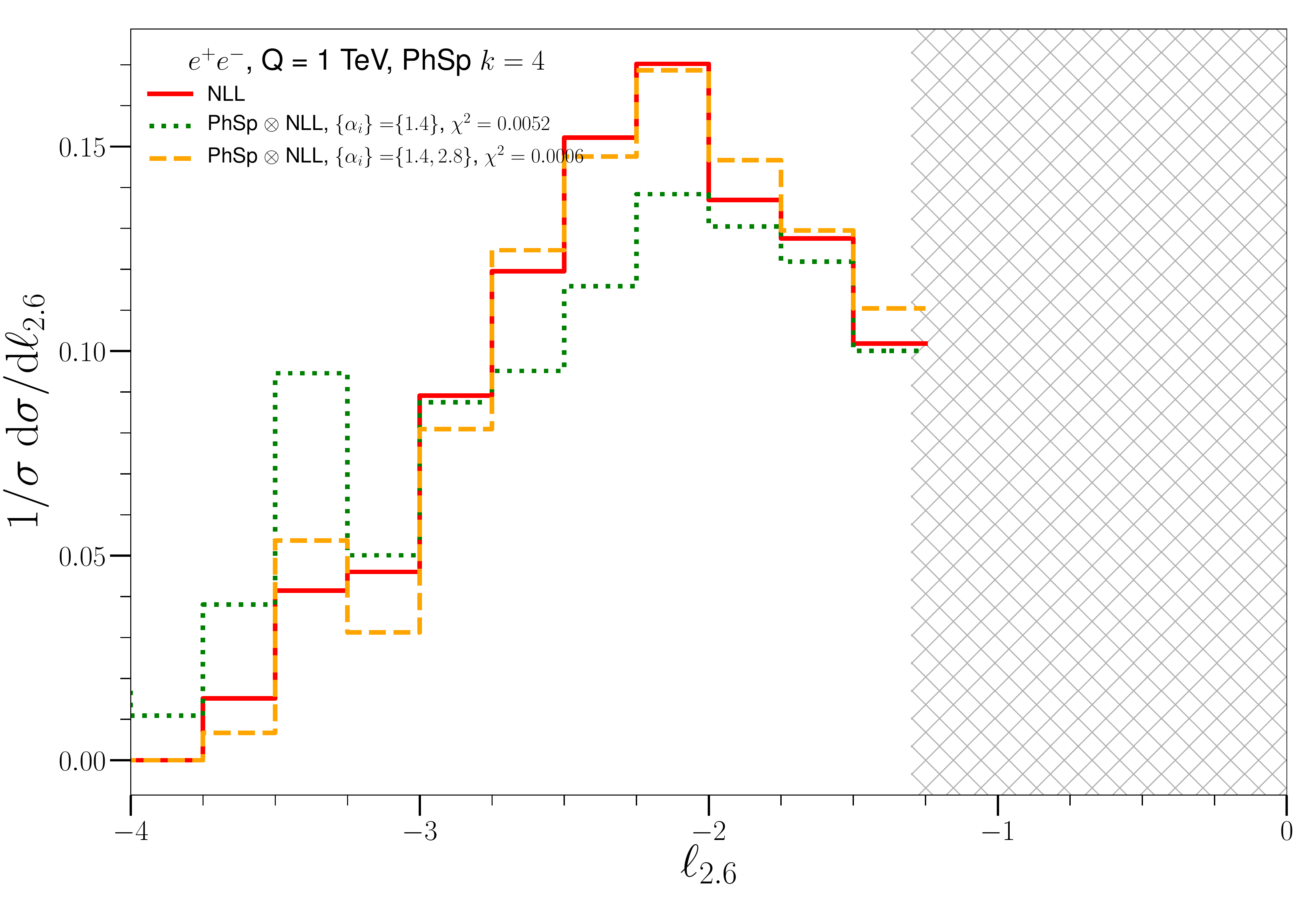}
\caption{Same as \fig{single}, but using our NLL analytic predictions instead of \Herwig.}
\label{fig:singleresum}
\end{figure}

\Fig{singleresum} shows the corresponding set of plots constructed by reweighing the NLL resummed results obtained from the calculation in \sec{jointresum}. In this case we have restricted the set of angularities that we examine to $\alpha_i = 0.2\times s$ with $s = 6, 7, \dots, 15$ instead, and we have used the ``best set'' of one or two angularities obtained from the analogous procedure done with \Herwig.
The restriction on the angularity exponents that we consider is chosen such that it allows for a sufficient number of bins in the analytical resummation to be populated, which is otherwise not the case for lower values of $\alpha_i$.
To focus solely on the differences that arise due to reweighing with a different number of angularities, all the distributions that enter in these plots are obtained from projecting the full three-dimensional distribution with angularity exponents $\{\alpha_j, 1.4, 2.8\}$. 
The reason for this is that the projection of an analytically resummed cross section involving a higher number of angularities down to a cross section involving a lower number of angularities does not exactly agree with the corresponding cross section obtained from a direct analytic calculation, i.e.~without any projection. A more in-depth discussion is relegated to \app{extra}. With these comments in mind, we note that the reweighed results of \fig{singleresum} show a similar trend as those of \fig{single}, constructed using \Herwig distributions.

\begin{figure}[t]
\centering
\includegraphics[width=0.45\textwidth]{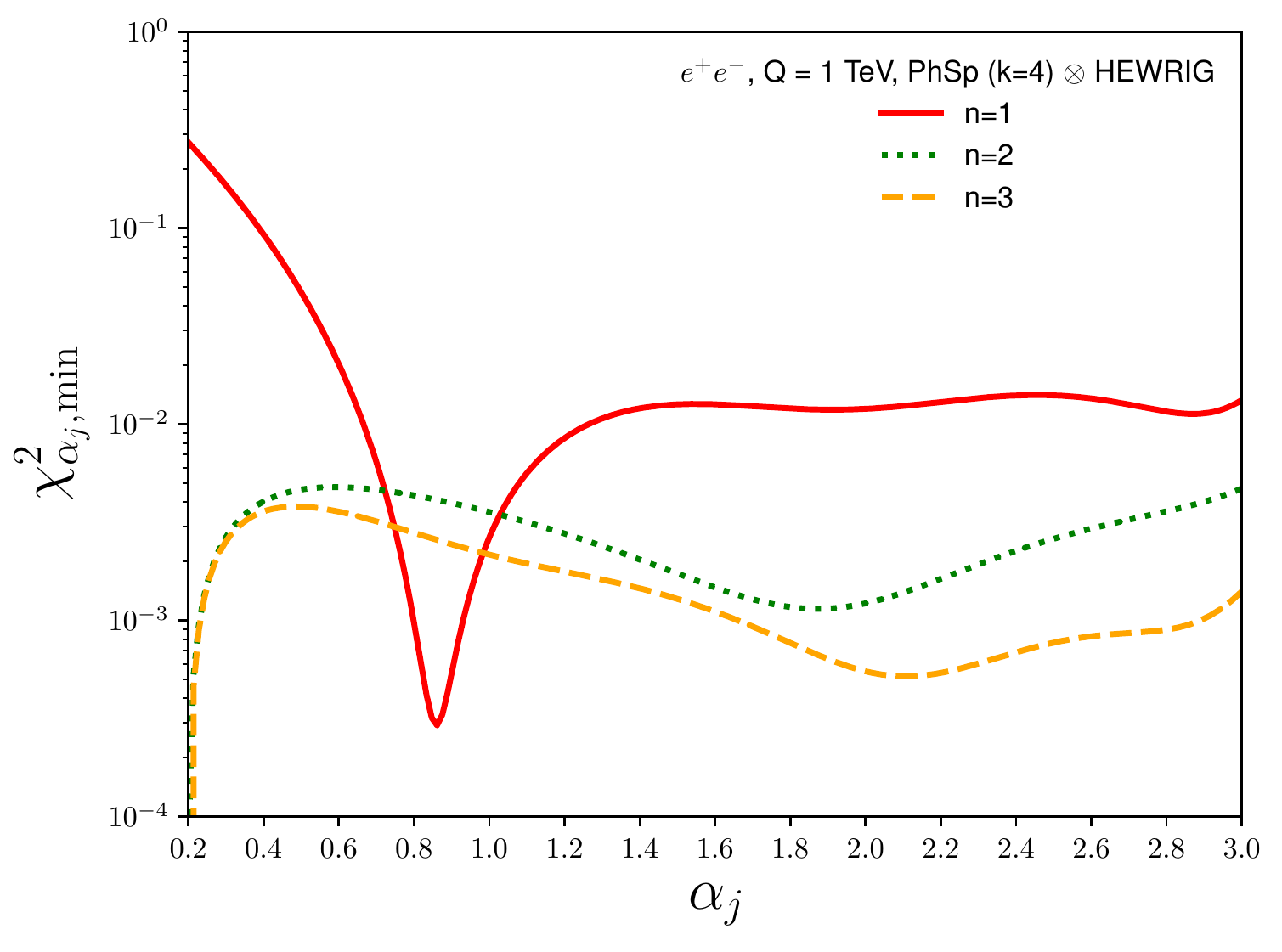}%
\includegraphics[width=0.45\textwidth]{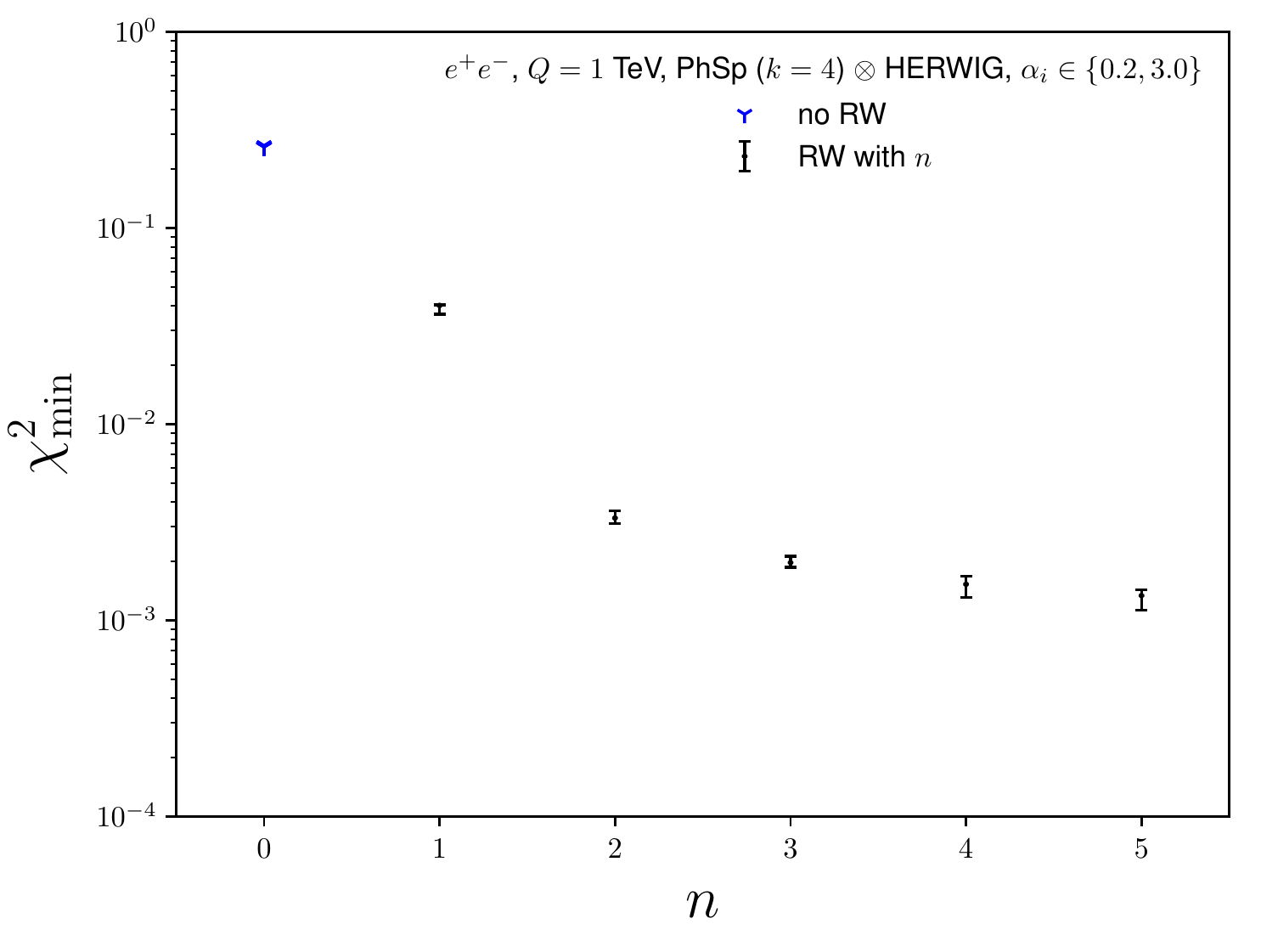}%

\caption{Left panel: We show the goodness-of-fit $\chi_{\alpha_j}^2$ from \eq{chi2_beta} for the best set of $n=1$ (red), $n=2$ (green dotted) and $n=3$ (yellow dashed) reweighed angularities as a function of the angularity exponent $\alpha_j$ for \Herwig. For each point we take the median value of the 11 replicas. Right panel: The global minimum $\chi^2_{\rm min}$ from \eq{chi2_total} for $n=1, \dots, 5$. The error bars represent the uncertainty as described in the text, with the central black dot representing the median over the replicas. The initial value of $\chi^2$ (blue star) shows how well flat phase space reproduces the angularities prior to reweighing.} 
\label{fig:global}
\end{figure}

To indicate the improvement obtained over the full domain of considered angularities, we show the goodness-of-fit, $\chi_{\alpha_j, {\rm min}}^2$ for the best set of $n=1,2,3$ reweighed angularities as a function of $\alpha_j$ in the left panel of \fig{global}. 
For $n=1$, one would expect that $\chi_{\alpha_j, {\rm min}}^2$ goes to zero for the best single angularity. However, we take the median of 11 replicas and the optimal single angularity is not the same for each of these. On the other hand, $\al_i = 0.2$ is always part of the set of best angularities for $n=2$. It is clear that the $n=2$ case performs substantially better than $n=1$, though there are a few angularities for which $n=2$ performs worse. This happens because they are close to the best single angularity and therefore reproduced very well by $n=1$, but not as well for $n=2$ since the best two angularities are further away. For $n=3$, there is a non-negligible, but less significant, improvement over $n=2$. 

Finally, in the right panel of \fig{global}, we show the minimum goodness-of-fit $\chi^2_{\rm min}$ for $n=1, \dots, 5$. For $n=1,2,3$ this is the global minimum, whereas for $n=4, 5$ the results were obtained iteratively, as described in \sec{reweigh}. For comparison we include $n=0$, which simply states how well pure phase-space predictions (without any reweighing) describe the angularity distributions in \Herwig, thus providing a baseline. 
The black error bars correspond to roughly one standard deviation, having been constructed from the spread of the 7 most central replicas out of a produced total of 11 replicas. The dots represent the median of the replicas. 
 One can again observe a substantial improvement from $n=1$ to $n=2$ and a smaller but still visible improvement from $n=2$ to $n=3$. The degree of improvement for going to $n=4$ or $n=5$ reweighed angularities is much smaller.  

In \app{robustness} we provide plots that demonstrate the robustness of our results under different variations. These include the use of \Pythia as the Monte Carlo for the reweighing, considering either 5- or 6-body phase space, restricting the values of angularity exponents, and considering lower or higher center-of-mass energies.
We also show results that discuss the quality of the projections from higher-dimensional distributions and the impact of these on the reweighing procedure in \app{projections}.

%%%%%%%%%%%%%%%%%%%%%%%%%%%%%%%%%%%%%%%%%%%%%%%%%%%%%%%%%%%%%%%%%%%%%%%%%%%%%%%%
\section{Conclusions}
\label{sec:conc}
%%%%%%%%%%%%%%%%%%%%%%%%%%%%%%%%%%%%%%%%%%%%%%%%%%%%%%%%%%%%%%%%%%%%%%%%%%%%%%%%

We have investigated the benefits and limitations of joint resummation of large logarithms, using as an example the resummation of $n$ angularities in $e^+ e^-$ collisions. A major part of this work involved the development of an analytical method to jointly resum, at next-to-leading logarithmic order, any number of angularities.  This was achieved using factorization theorems derived in the SCET formalism. While the joint resummation of two angularities had been studied before~\cite{Larkoski:2014tva, Procura:2014cba, Procura:2018zpn}, identifying all the regimes and relevant modes, and estimating the power corrections to connect them becomes more complicated for three (or more) angularities. This can be extended to processes with jets in hadronic collisions, in which case gluon jets also enter, and non-global logarithms~\cite{Dasgupta:2001sh} arise from soft radiation that simultaneously contributes to the angularities (measured on the jet) and the out-of-jet region.

Taking distributions obtained from this analytical resummation, as well as from the \Herwig and \Pythia Monte Carlo parton showers, we have studied whether employing them to reweigh a flat phase-space generator leads to improved predictions for other angularities (not used as input) via kinematic correlations. We have found an order of magnitude improvement when reweighing by distributions of two angularities over using only one, demonstrating the benefit of joint resummation. Reweighing with three or more angularities provides further improvement, albeit with a diminishing effect. The robustness of our conclusions is demonstrated by varying parts of our setup.

Our study shows that reweighing leads to improved predictions, particularly if the observable used in the reweighing procedure is similar to the observable of interest. Augmenting Monte Carlo parton showers by analytic resummation at NLL is probably not that useful, due to the sizable perturbative uncertainty at this order. However, this could be improved by matching the NLL to a fixed-order calculation. Furthermore, the factorization formulae presented here are not limited to a specific resummation order, and in principle all ingredients needed to NNLL are the same as for two angularities in ref.~\cite{Procura:2018zpn}. The reason for this is that, apart from the anomalous dimensions, all  ingredients in the factorization formulae are only needed at one-loop order. They can therefore depend on at most two independent variables, so any additional angularities can be expressed in those two. 

We believe that this approach opens up a new route towards precise and detailed (i.e.~differential) predictions for collisions at the LHC, supplementing current advances in Monte Carlo parton showers. 
Such predictions are particularly important in an era in which the Standard Model is subjected to ever more stringent tests, and Machine Learning techniques are developed in order to uncover faint signals through detailed features in the data.

%%%%%%%%%%%%%%%%%%%%%%%%%%%%%%%%%%%%%%%%%%%%%%%%%%%%%%%%%%%%%%%%%%%%%%%%%%%%%%%%
\begin{acknowledgments}
This work was supported by the ERC grant ERC-STG-2015-677323, and  the D-ITP consortium, a program of the Netherlands Organization for Scientific Research (NWO) that is funded by the Dutch Ministry of Education, Culture and Science (OCW). This article  is based upon work from COST Action CA16201 PARTICLEFACE, supported by COST (European Cooperation in Science and Technology).
\end{acknowledgments}
%%%%%%%%%%%%%%%%%%%%%%%%%%%%%%%%%%%%%%%%%%%%%%%%%%%%%%%%%%%%%%%%%%%%%%%%%%%%%%%%

\appendix

%%%%%%%%%%%%%%%%%%%%%%%%%%%%%%%%%%%%%%%%%%%%%%%%%%%%%%%%%%%%%%%%%%%%%%%%%%%%%%%%
\section{Additional plots}
\label{app:extra}
%%%%%%%%%%%%%%%%%%%%%%%%%%%%%%%%%%%%%%%%%%%%%%%%%%%%%%%%%%%%%%%%%%%%%%%%%%%%%%%%

In this appendix we provide some additional plots that strengthen the universality of our conclusions, and briefly discuss the projections of the analytical resummed results.

%===============================================================================
\subsection{Robustness of conclusions}\label{app:robustness}
%===============================================================================

\begin{figure}[t]
\centering
\includegraphics[width=0.45\textwidth]{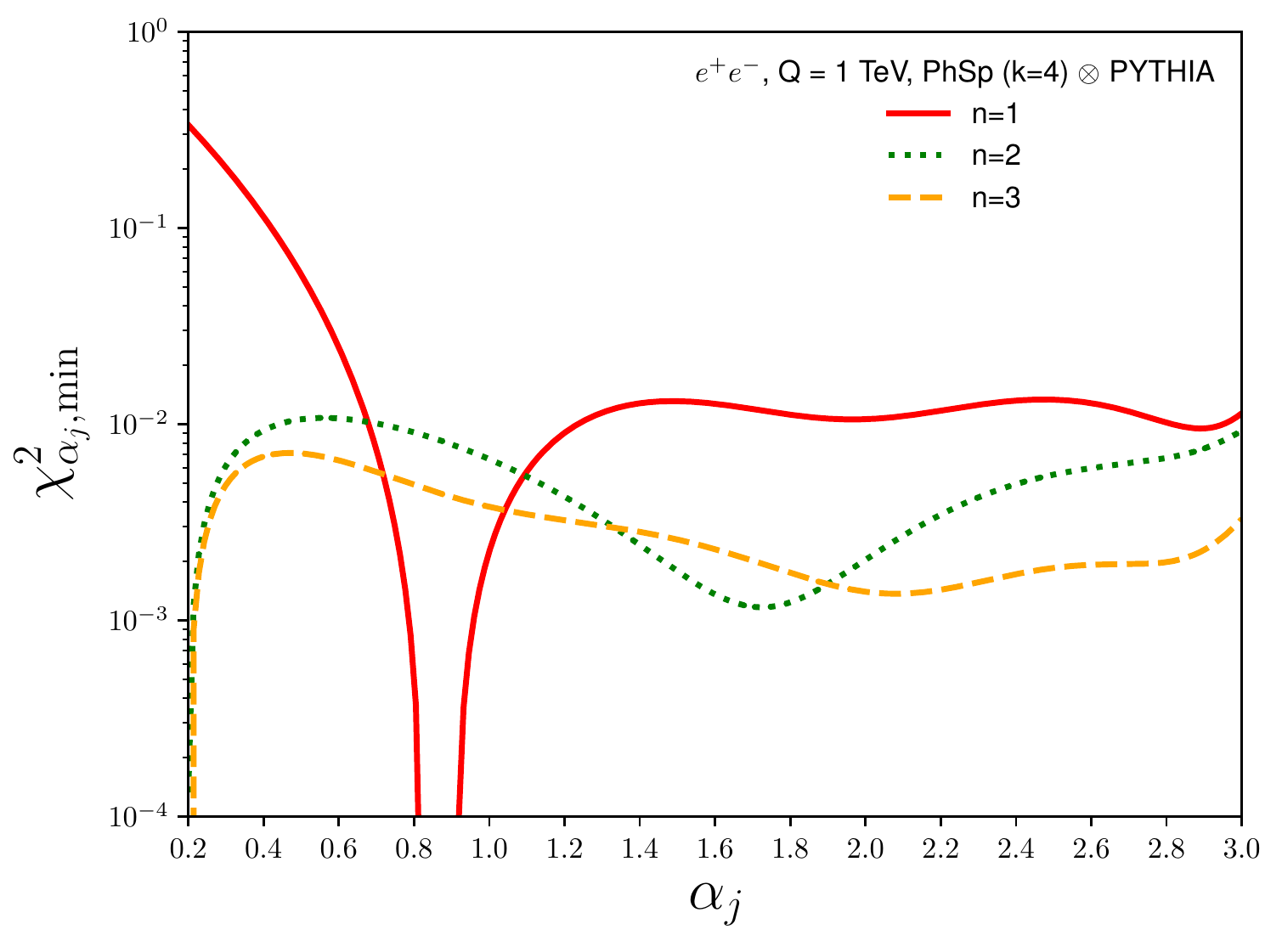}%
\includegraphics[width=0.45\textwidth]{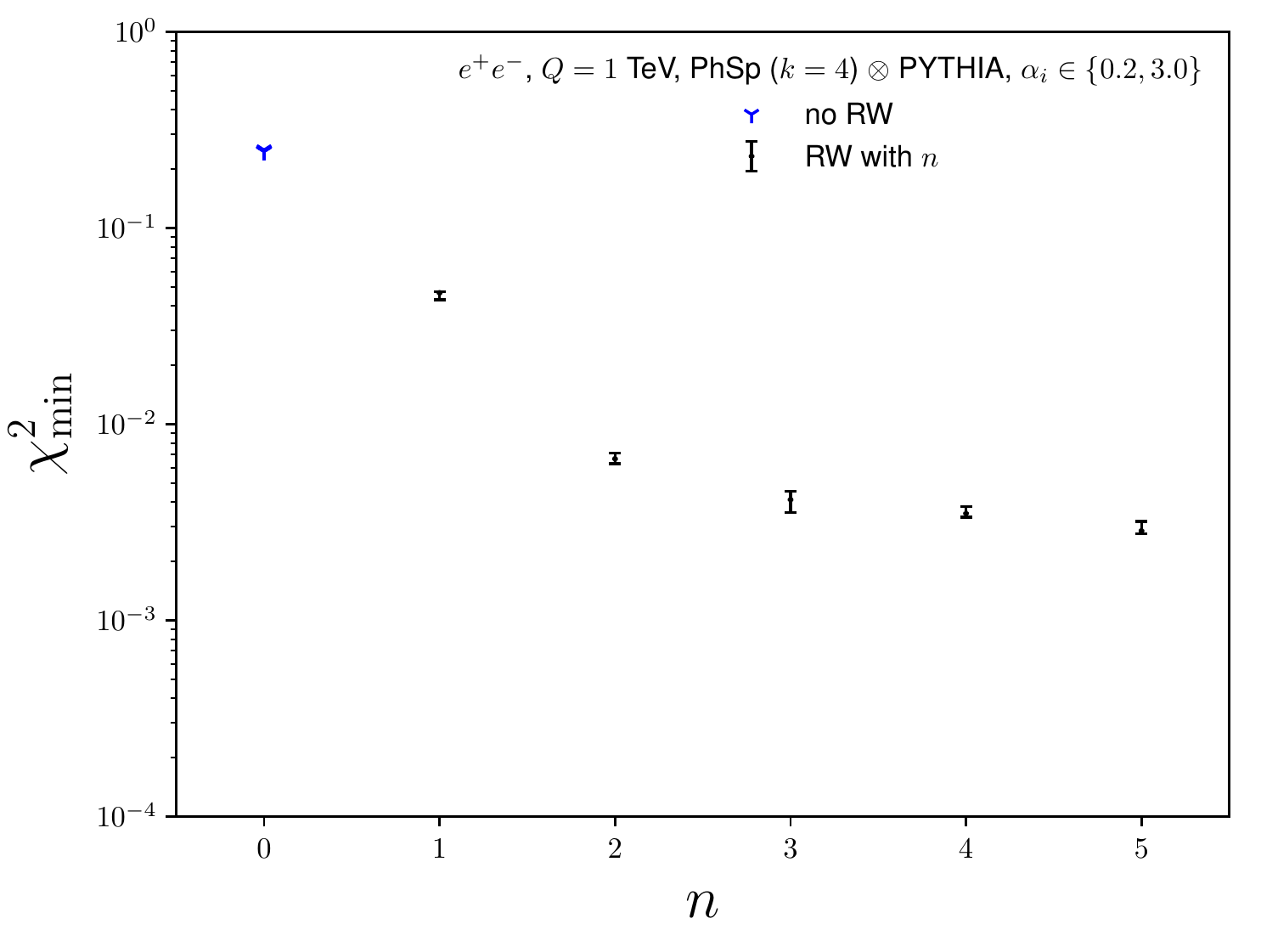}%
\caption{Same as \fig{global} but using \Pythia.} 
\label{fig:globalpythia}
\end{figure}

First of all, we show the analogue of \fig{global} using \Pythia instead of \Herwig. Specifically, the left panel of \fig{globalpythia} shows the goodness-of-fit, $\chi_{\alpha_i, {\rm min}}^2$, as a function of $\alpha_i$. In this case, there is a single best angularity at $\al=0.8$ for each of the replicas. Also, there is now a small region that does not improve from $n=2$ to $n=3$. The right panel shows the global minimum goodness-of-fit $\chi^2_{\rm min}$ for $n=1, \dots, 5$. The qualitative behavior is very similar to that in the right panel of \fig{global} for \Herwig, but the values of $\chi_{\alpha_i, {\rm min}}^2$ are somewhat larger.

\begin{figure}[t]
\centering
\includegraphics[width=0.45\textwidth]{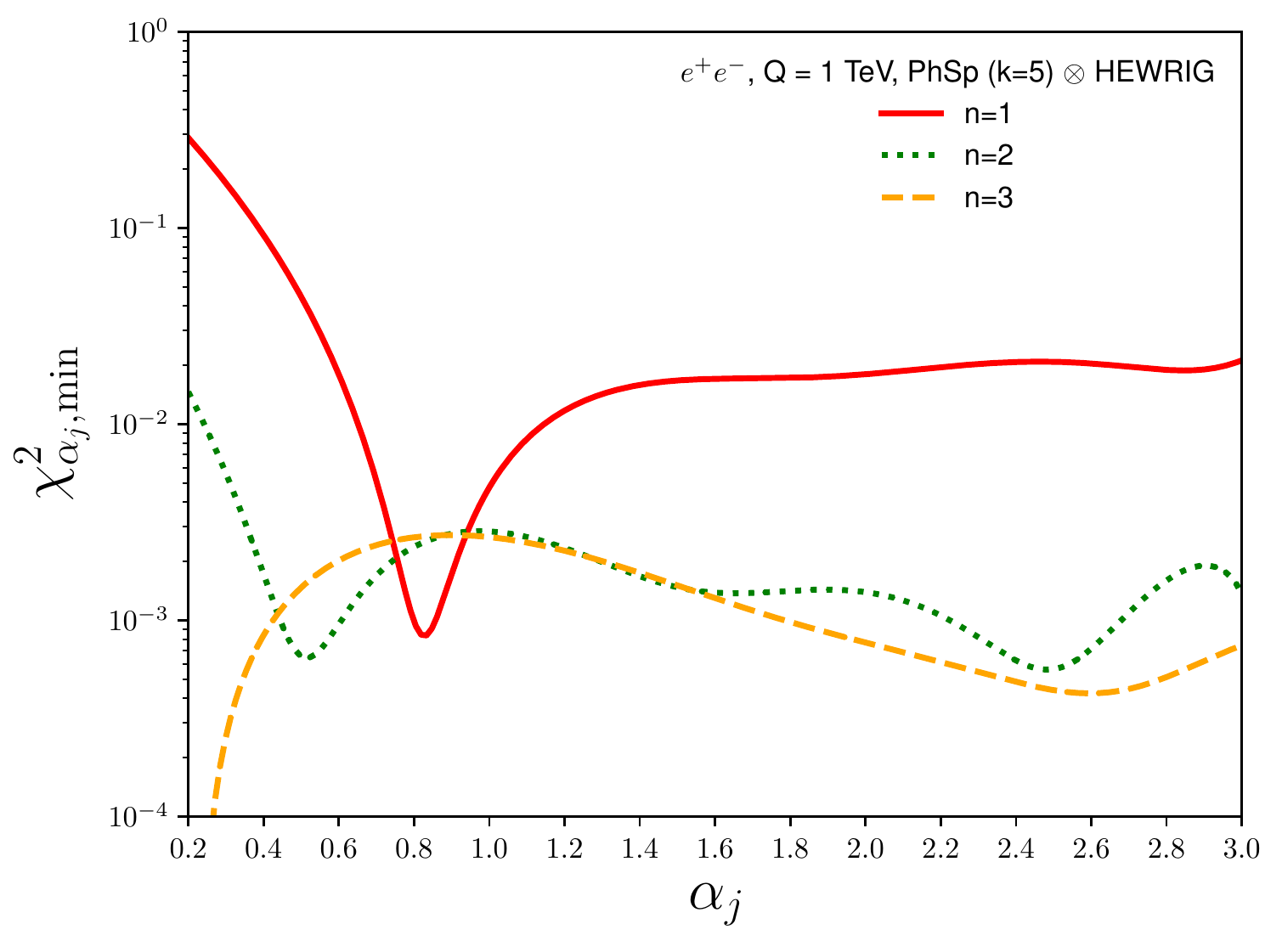}%
\includegraphics[width=0.45\textwidth]{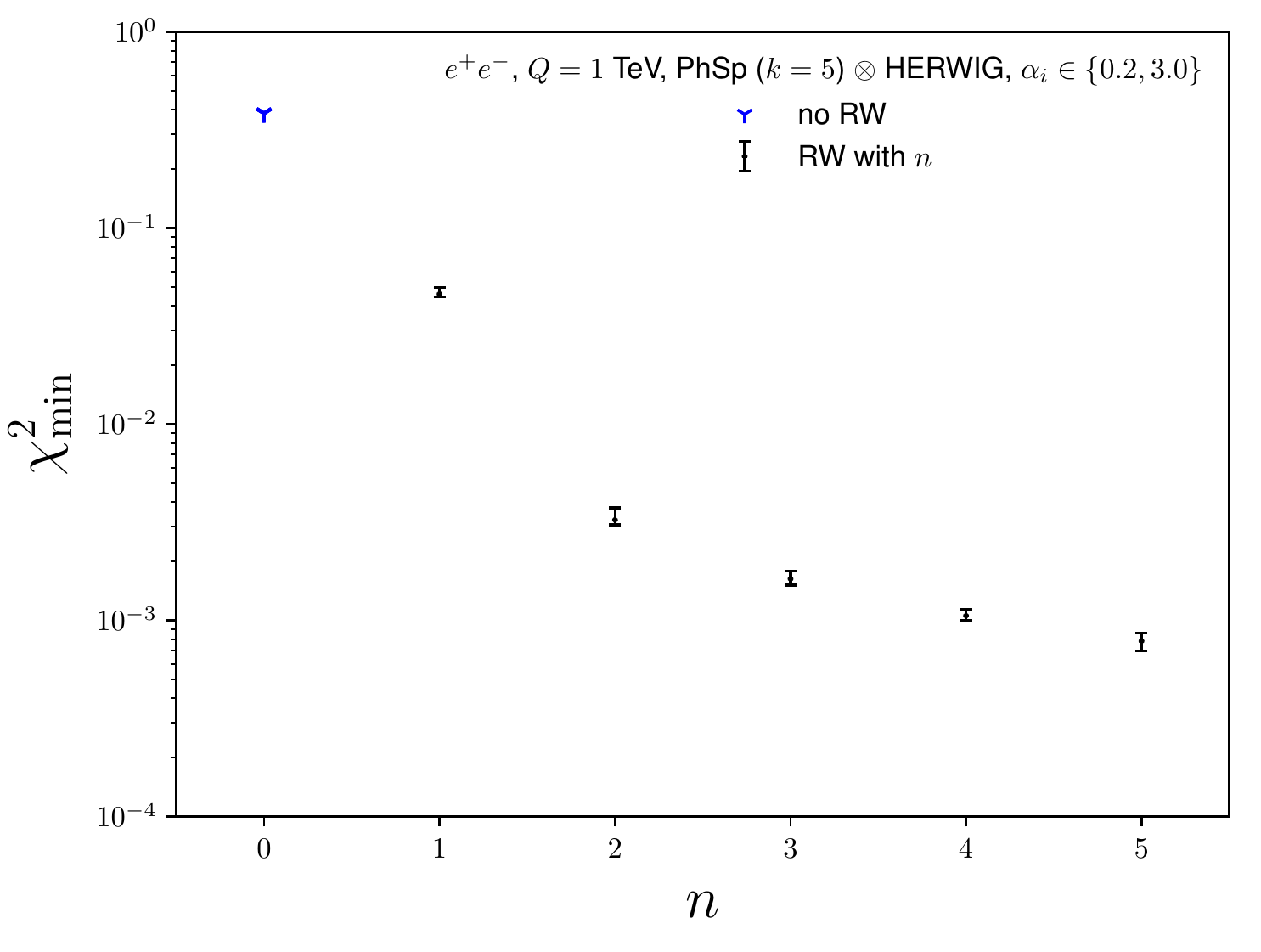}%
\caption{Same as \fig{global} but using $5$-body phase space in the reweighing.} 
\label{fig:globalk5}
\end{figure}

\Fig{globalk5} is the analogue of \fig{global}, but using $5$-body flat phase space (i.e.~$k=5$) instead of $4$-body phase space. While the qualitative behavior is largely the same, there are some numerical differences that are mostly driven by statistical fluctuations, most visible in the left panel. Specifically, the sampling of the collinear and soft regions of phase space relevant for angularities is worse for $k=5$. 
 
\begin{figure}[t]
\centering
\includegraphics[width=0.45\textwidth]{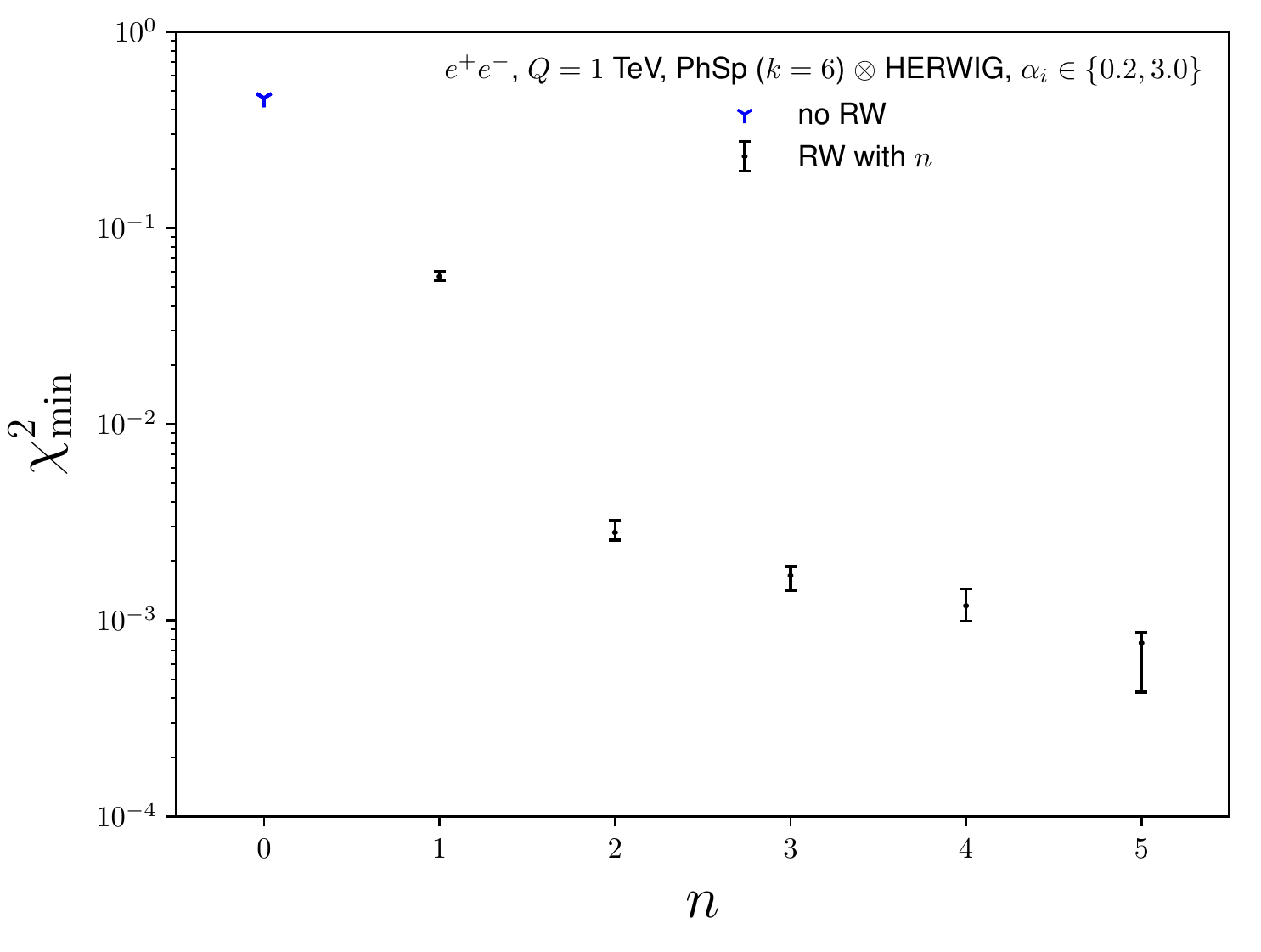}%
\includegraphics[width=0.45\textwidth]{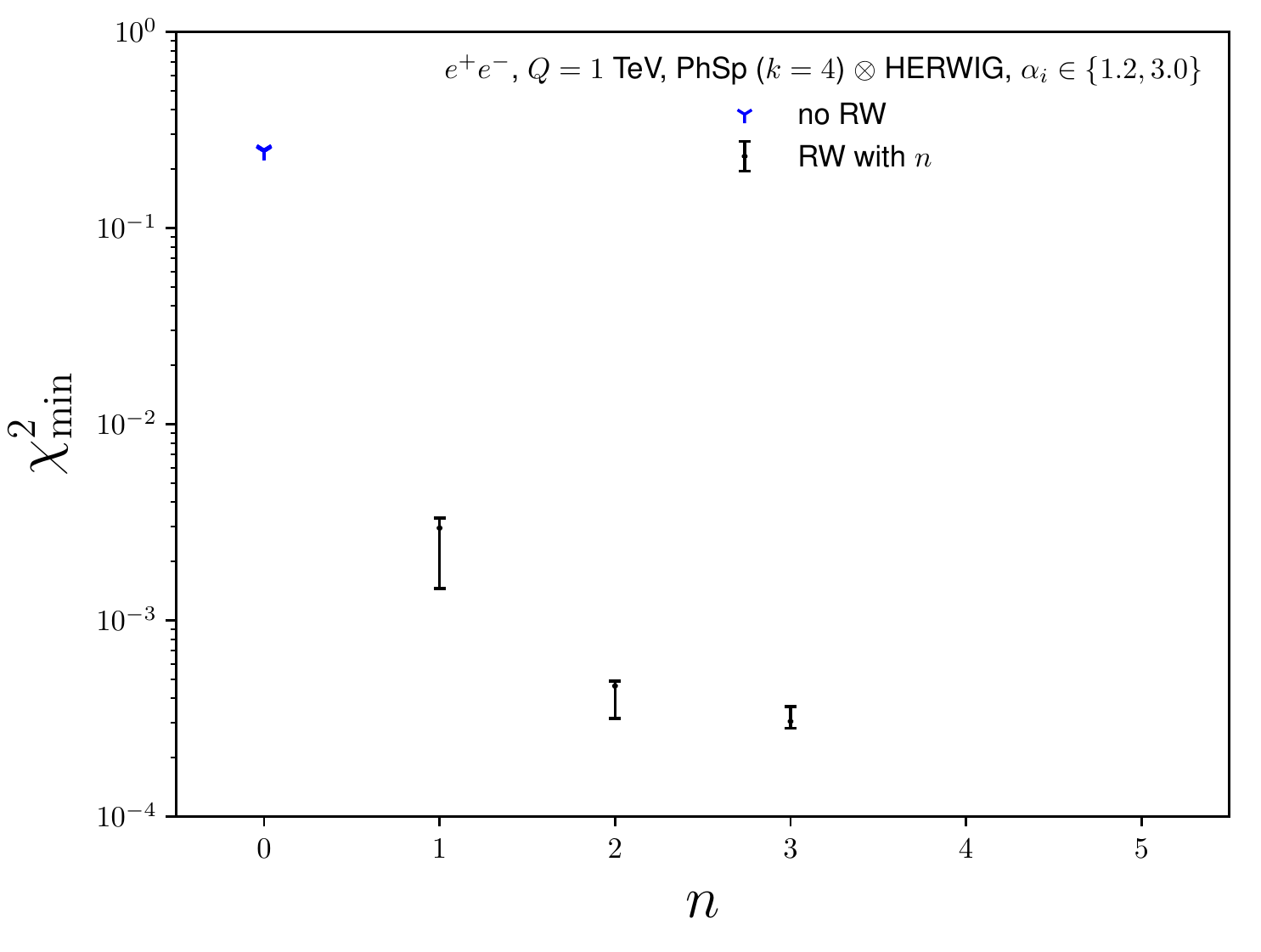}%
\caption{Same as the right panel of \fig{global}, but obtained from $k=6$-body phase space (left panel) or by restricting the domain of angularity exponents to the set $\alpha_i \in [1.2, 3.0]$ (right panel).} 
\label{fig:globalk6_restricted}
\end{figure}

The left panel of \fig{globalk6_restricted} is the analogue of the right panel of \fig{global}, showing the minimum goodness-of-fit $\chi^2_{\rm min}$ for $6$-body phase space. As with $k=5$, the larger number of phase-space particles worsens the sampling of the angularity phase space and hence increases the statistical fluctuations, reflected by the larger uncertainties. The right panel shows $\chi^2_{\rm min}$ where we repeated the analysis for $k=4$, restricting to the angularity exponents in the interval $\alpha_i \in [1.2, 3.0]$ (again in steps of $0.2$). Because the total number of angularities is smaller, we only show the result of reweighing with $n=1,2,3$ angularities. This is also the reason for the faster convergence as function of the number $n$ of angularities used in the reweighing.

\begin{figure}[t]
\centering
\includegraphics[width=0.45\textwidth]{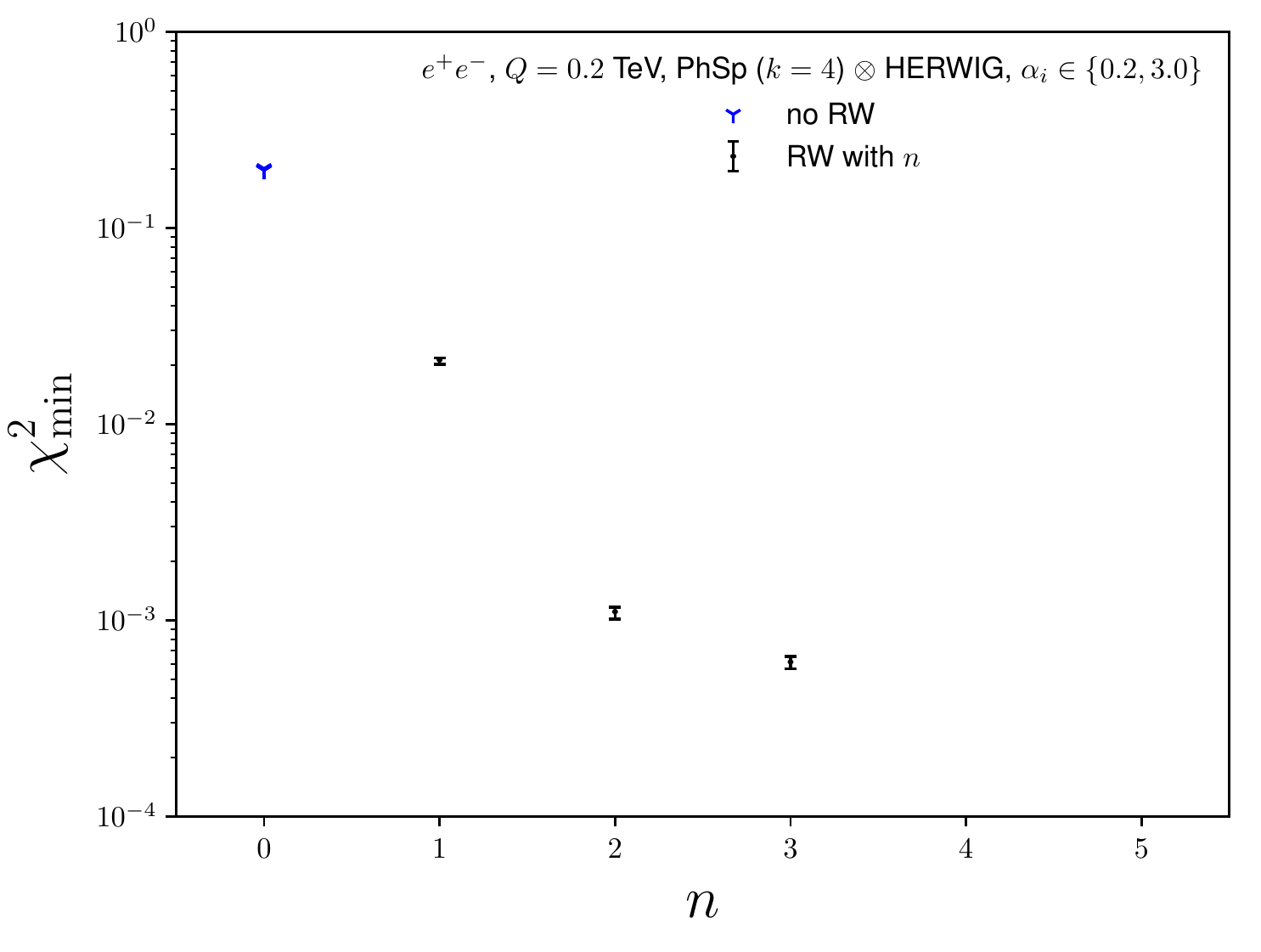}%
\includegraphics[width=0.45\textwidth]{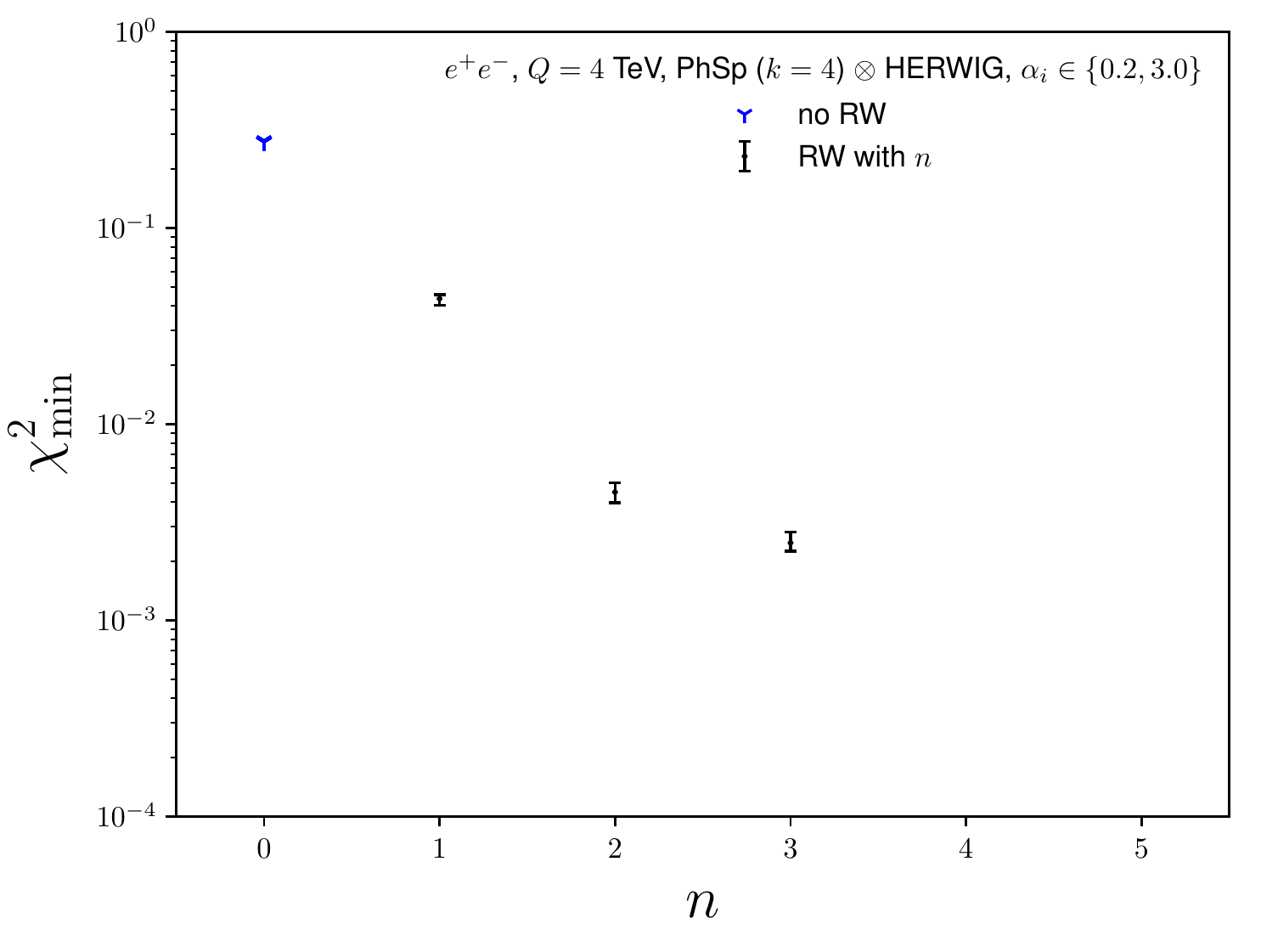}%
\caption{Same as the right panel of \fig{global} but for $Q=200$~GeV (left panel) or $Q=4$~TeV (right panel). In this case, $n=4,5$ are not shown.} 
\label{fig:globalQ}
\end{figure}

Finally, \fig{globalQ} is the analogue of the right panel of \fig{global}, but for different center-of-mass energies, $Q=200$~GeV (left panel) and $Q=4$~TeV (right panel). We do not show the reweighing with $n=4,5$ angularities. The qualitative behavior is similar as for $Q=1$~TeV, but suggests that the reweighing procedure performs better for lower energies. This is in line with what one would expect from the increase in jet entropy with $Q$~\cite{Neill:2018uqw}.

%===============================================================================
\subsection{Projections}\label{app:projections}
%===============================================================================

In \fig{singleresum} we showed results obtained by applying the reweighing procedure to analytical resummed results, all derived from the same multi-dimensional calculation. Here we comment briefly on the issues leading to this approach, and show results for an alternative method.

\begin{figure}[t]
\centering
\includegraphics[width=0.45\textwidth]{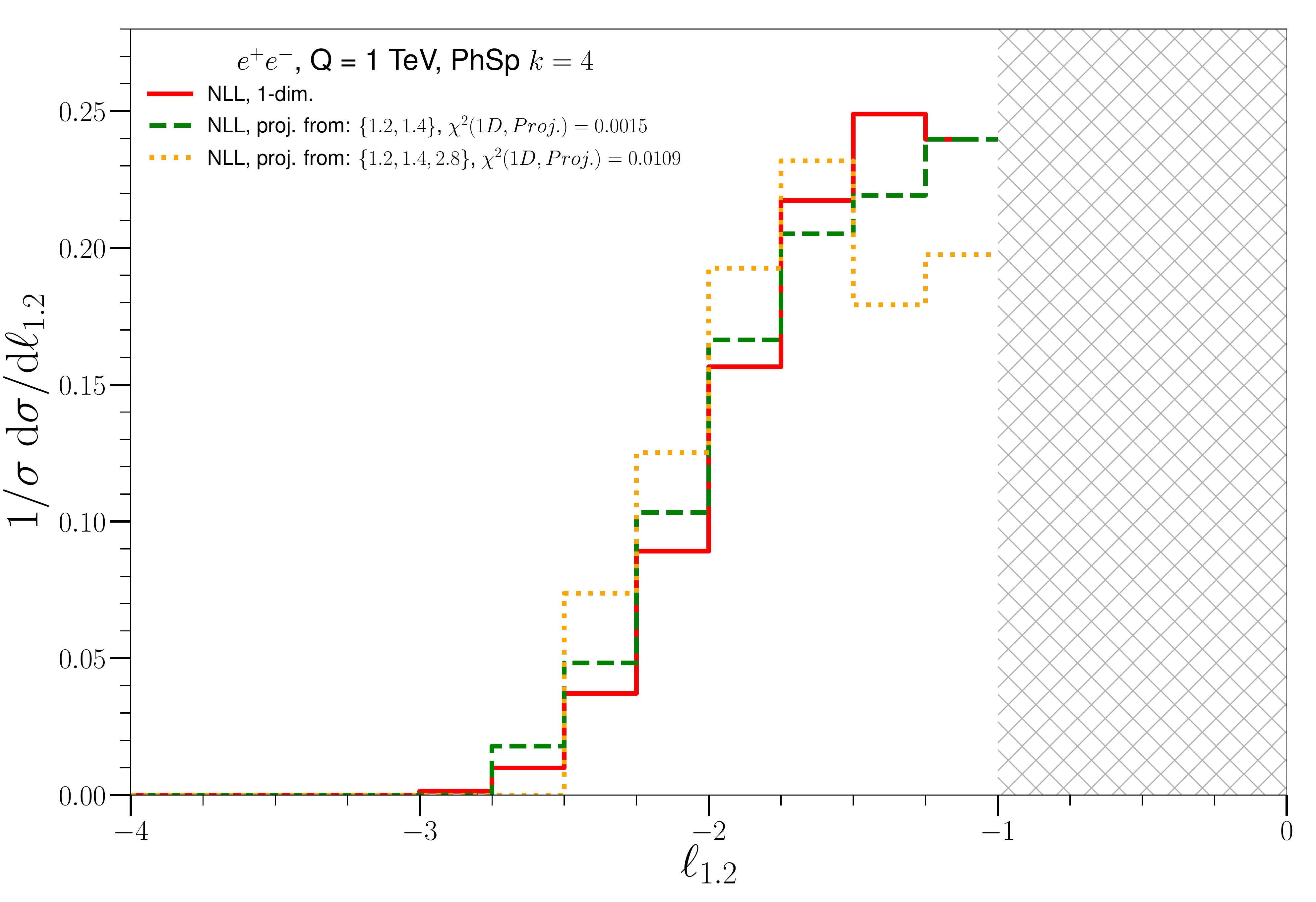}%
\includegraphics[width=0.45\textwidth]{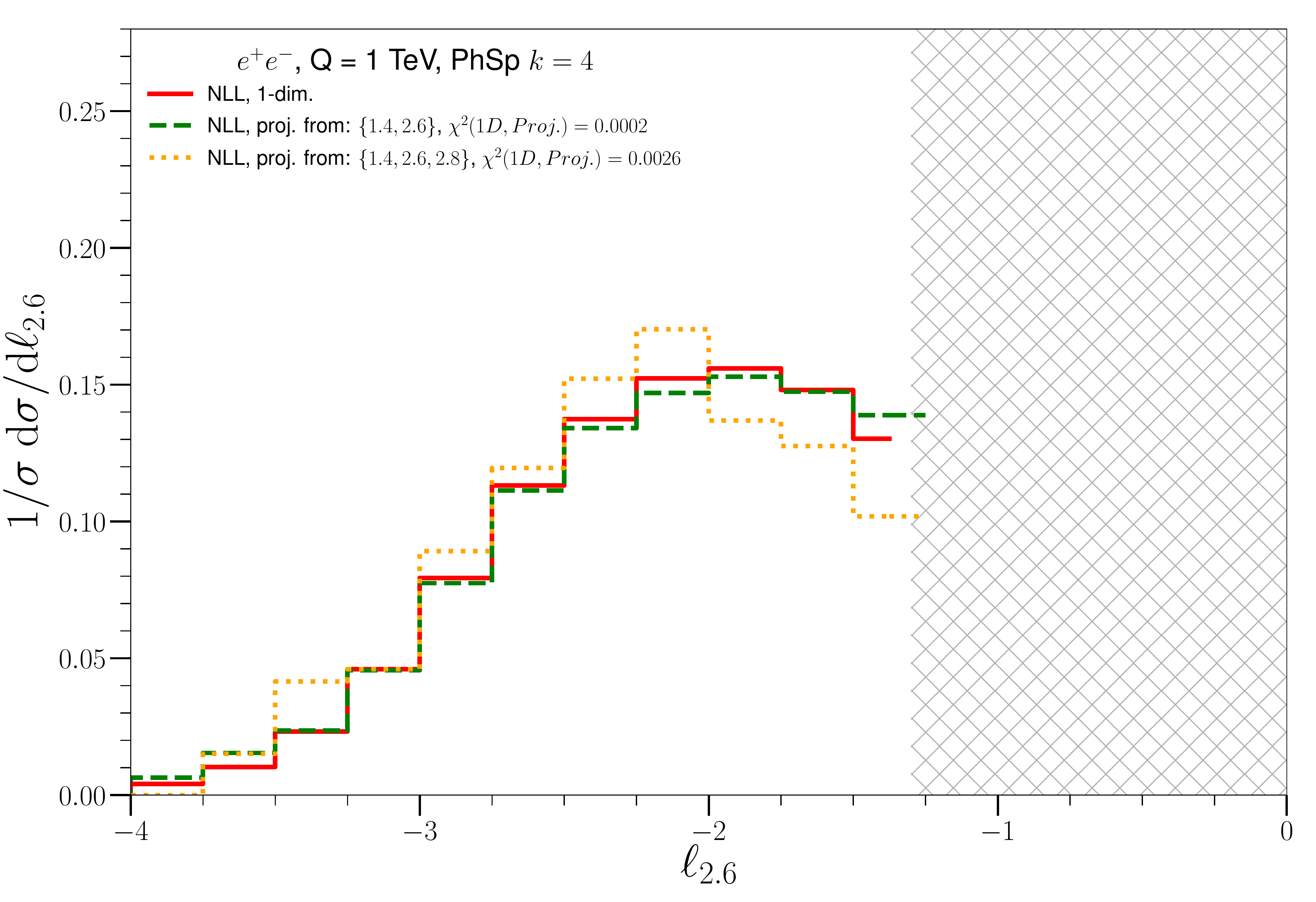}
\caption{The projections from multi-differential angularity distributions down to the distribution for a single angularity for two specific choices $\alpha_j = 1.2$ (left panel) and 2.6 (right panel). The $\chi^2$ gives the difference between the one-dimensional resummed angularity distribution and each of these projections. }
\label{fig:projections}
\end{figure}

\begin{figure}[t]
\centering
\includegraphics[width=0.45\textwidth]{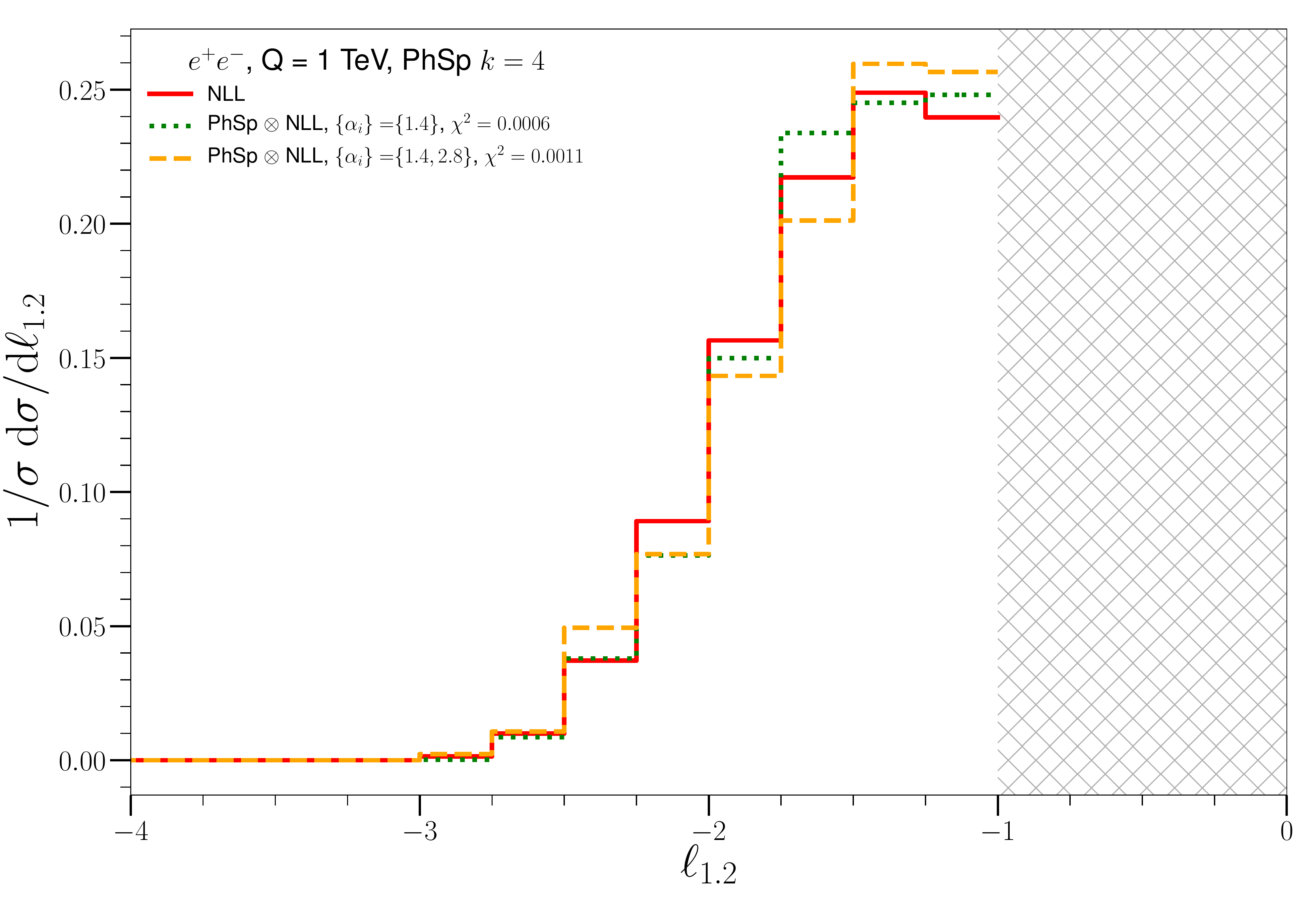}%
\includegraphics[width=0.45\textwidth]{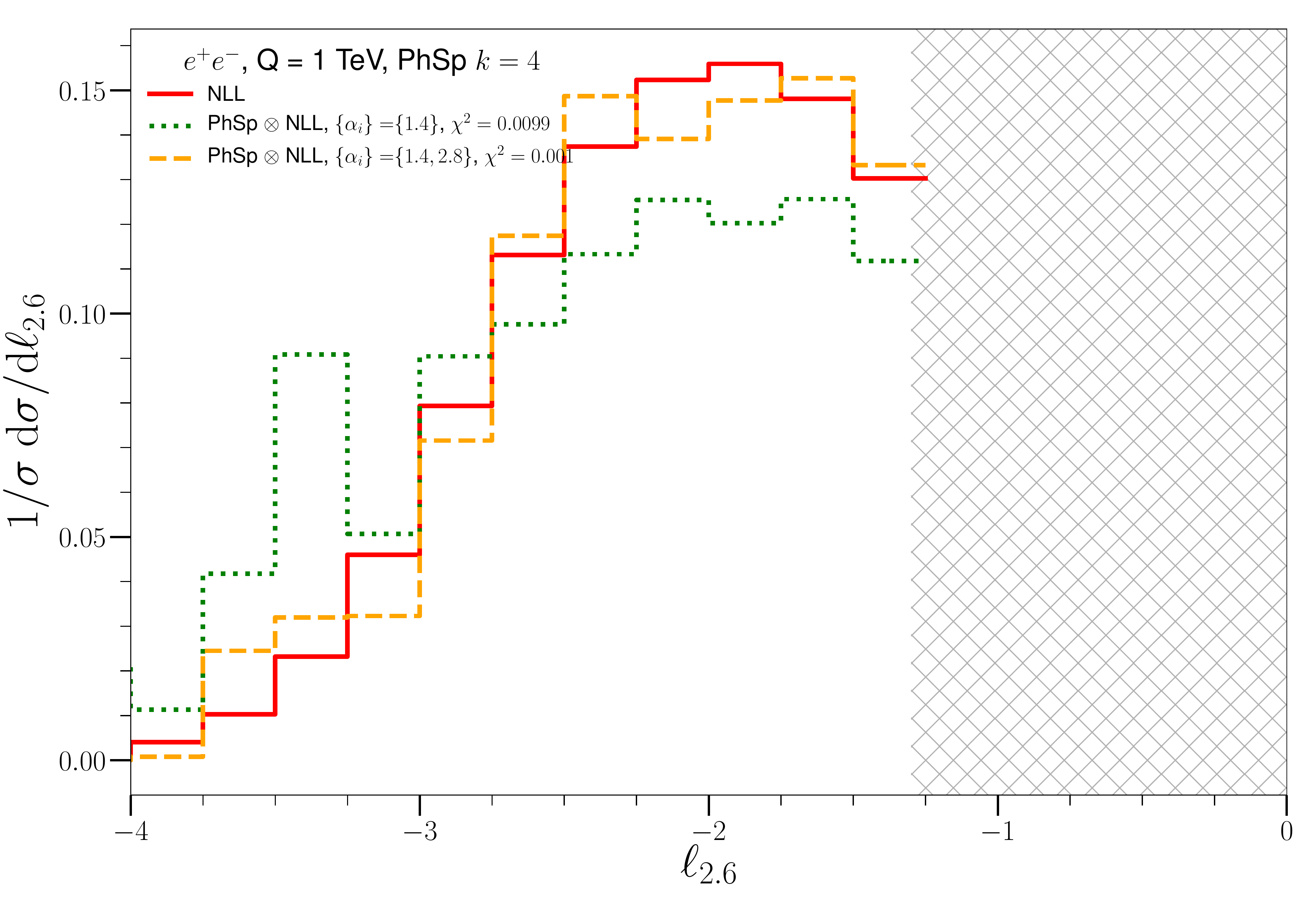}
\caption{Same as \fig{singleresum}, but with the reweighing performed using the multi-dimensional analytical distributions calculated directly (rather than projections from a single multi-dimensional distribution).}
\label{fig:singleresum_notproj}
\end{figure}

\Fig{projections} shows the projections from higher-dimensional analytical results down to two specific angularity choices $\alpha_j = 1.2$ and $\alpha_j = 2.6$. The yellow dotted and green dashed curves represent the projection from the indicated three-dimensional and two-dimensional cross section respectively. The red curve shows the directly determined cross section differential in the single angularity. The figure clearly illustrates that the curves are not identical and that projections differ more in comparison to the one-dimensional distribution if one starts from higher-dimensional cross sections. This fact is quantified in the $\chi^2$ between the one-dimensional resummed angularity distribution and each of the projections. Note that there is no corresponding issue for the \Herwig and \Pythia results, since they originate from fully exclusive events. 

We suspect that this discrepancy is largely due to binning issues. As described in \subsec{Lund}, the number of distinct kinematic regions in phase space increases dramatically when cross sections differential in more angularities are considered. The result of the differential cross section in (the center of) each bin is obtained by determining the cumulative cross section on the edges of the bin\footnote{The use of cumulative cross sections is required to, at least theoretically, enable the recovery of the inclusive cross section~\cite{Abbate:2010xh, Almeida:2014uva, Alioli:2015toa, Bertolini:2017eui}.} and taking a numerical derivative. Due to the relatively small number of bins and the increasing number of kinematic regions, situations in which the edges and the center of a bin lie in different kinematic regions might occur. In these cases, the prediction of the spectrum (at the center of the bin) is obtained from input provided by cumulative distributions obtained from factorization formulas that are not valid at that point. As this is a binning issue, we expect the effect to diminish when a larger number of bins is considered.

 For completeness, we show in \fig{singleresum_notproj} the reweighing performed for the same restricted set of angularities as in \fig{singleresum}, but using the $n$-dimensional analytical distributions calculated directly, and comparing to the one-dimensional analytic prediction. 

%%%%%%%%%%%%%%%%%%%%%%%%%%%%%%%%%%%%%%%%%%%%%%%%%%%%%%%%%%%%%%%%%%%%%%%%%%%%%%%%
\section{Resummation}
\label{app:resum}
%%%%%%%%%%%%%%%%%%%%%%%%%%%%%%%%%%%%%%%%%%%%%%%%%%%%%%%%%%%%%%%%%%%%%%%%%%%%%%%%

The perturbative functions occurring in the factorization formula in \eq{factorization_bulk_two} are renormalized through
%%%
\begin{align} \label{eq:RGEs}
\mu\frac{\df}{\df \mu} H(Q^2,\mu) &= \gamma_H(Q^2,\mu)\, H(Q^2,\mu)\, ,\nn\\
\mu\frac{\df}{\df \mu} J(Q^\al e_\al,\mu) &= \gamma_J(Q^\al e_\al,\mu) \underset{\al}{\otimes} J(Q^\al e_\al,\mu)\, ,\nn\\
\mu\frac{\df}{\df \mu} S(Q^\al e_\al,\mu) &= \gamma_S(Q^\al e_\al,\mu) \underset{\al}{\otimes} S(Q^\al e_\al,\mu)\, ,\nn\\
\mu\frac{\df}{\df \mu} \cS(Q^{\al_i} e_{\al_i},Q^{\al_j} e_{\al_j},\mu) &= \gamma_\cS(Q^{\al_i} e_{\al_i} ,Q^{\al_j} e_{\al_j},\mu) \underset{\al_i,\al_j}{\otimes} \cS(Q^{\al_i} e_{\al_i},Q^{\al_j} e_{\al_j},\mu)
\,.\end{align}
%%%
with anomalous dimensions given by
%%%
\begin{align} \label{eq:anomDims}
\gamma_H(Q^2,\mu) &= 2\Ga_\cusp(\al_s)\ln\Big(\frac{Q^2}{\mu^2}\Big) + \ga_H(\al_s)\, , \nn\\
\gamma_J(Q^\al e_\al,\mu) &= - \frac{2}{\al-1} \Ga_\cusp(\al_s) 
\frac{1}{\mu^\al}\,\cL_0\Big(\frac{Q^\al e_\al}{\mu^\al}\Big) + \ga_J(\al_s)\delta(Q^\al e_\al)\, , \nn\\
\gamma_S(Q^\al e_\al,\mu) &= \frac{4}{\al-1} \Ga_\cusp(\al_s) \frac{1}{\mu^\al}\,\cL_0\Big(\frac{Q^\al e_\al}{\mu^\al}\Big) \nn\\
& \quad + 
\Big[\ga_S(\al_s)-2 \Ga_\cusp(\al_s) \ln \Big(\frac{Q^2}{\mu^2}\Big)
\Big]\delta(Q^\al e_\al)\, , \nn\\
\gamma_\cS(Q^{\al_i} e_{\al_i},Q^{\al_j} e_{\al_j},\mu) &= - \frac{2}{\al_i-1} \Ga_\cusp(\al_s)\frac{1}{\mu^{\al_i}}\,\cL_0\Big(\frac{Q^{\al_i} e_{\al_i}}{\mu^{\al_i}}\Big)\de(Q^{\al_j}e_{\al_j}) \nn\\
& \quad + \frac{2}{\al_j - 1} \Ga_\cusp(\al_s) \frac{1}{\mu^{\al_j}}\,\cL_0\Big(\frac{Q^{\al_j} e_{\al_j}}{\mu^{\al_j}}\Big) \de(Q^{\al_i} e_{\al_i}) \nn\\
& \quad + \ga_\cS(\al_s)  \de(Q^{\al_i} e_{\al_i}) \de(Q^{\al_j}e_{\al_j})
\,.\end{align}
%%%
The $\overline{\mathrm{MS}}$ cusp anomalous dimension to two loops is given by~\cite{Korchemsky:1987wg}
%%%
\begin{align}
\Gamma_{\text{cusp}}(\alpha_s) = \frac{\al_s}{4\pi}\, 4C_F + \Big(\frac{\al_s}{4\pi}\Big)^2
\,\frac{4}{3} C_F \big[ (4 - \pi^2) C_A + 5 \beta_0 \big]
\,.\end{align}
%%%
The one-loop non-cusp anomalous dimensions are given by 
%%%
\begin{align}
\ga_S = \ga_\cS = 0\,, \qquad \ga_H = -\frac{3\al_sC_F}{\pi}\,, \qquad \ga_J = \frac{3\al_sC_F}{2\pi}
\,.\end{align}
%%%
To NLL accuracy, the evolution kernels occurring in the resummed cumulative cross section in \eq{resummed_cumulative} are given by
%%%
\begin{align} \label{eq:Keta}
K_\Gamma(\mu, \mu_0) &= -\frac{\Gamma_0}{4\beta_0^2}\,
\biggl[ \frac{4\pi}{\alpha_s(\mu_0)}\, \Bigl(1 - \frac{1}{r} - \ln r\Bigr)
   + \biggl(\frac{\Gamma_1 }{\Gamma_0 } - \frac{\beta_1}{\beta_0}\biggr) (1-r+\ln r)
   + \frac{\beta_1}{2\beta_0} \ln^2 r \biggr]
\,, \nn\\
\eta_\Gamma(\mu, \mu_0) &=
 - \frac{\Gamma_0}{2\beta_0}\, \biggl[ \ln r
 + \frac{\alpha_s(\mu_0)}{4\pi}\, \biggl(\frac{\Gamma_1 }{\Gamma_0 }
 - \frac{\beta_1}{\beta_0}\biggr)(r\!-\!1)
    \biggr]
\,, \nn\\
K_{\gamma_F}(\mu, \mu_0) &=
 - \frac{\gamma_{F,0}}{2\beta_0}\, \ln r
\,,\end{align}
%%%
where  $r = \alpha_s(\mu)/\alpha_s(\mu_0)$ has been defined. The required one- and two-loop coefficients of the beta function in the \MSbar~scheme are given by~\cite{Tarasov:1980au}
%%%
\begin{align}
\beta_0 &= \frac{11}{3}\,C_A -\frac{4}{3}\,T_F\,n_f\,, \qquad \beta_1 = \frac{34}{3}\,C_A^2  - \Bigl(\frac{20}{3}\,C_A\, + 4 C_F\Bigr)\, T_F\,n_f
\,.\end{align}
%%%

\addcontentsline{toc}{section}{References}
\bibliographystyle{jhep}
\bibliography{paper}

\end{document}